\documentclass[twocolumn]{aastex631}
\usepackage{CJKutf8} 
\usepackage{amsmath}

\begin{document}
\newcommand{\Dennis}[1]{{\bf\color{blue} DZ: #1}}
\newcommand{\RLD}[1]{{\it\color{blue} #1}}
\newcommand{\RD}[1]{{\color{blue}(RD: #1}}
\newcommand{\Huanian}[1]{{\bf\color{red} HZ: #1}}t
\newcommand{\vdag}{(v)^\dagger}
\newcommand\aastex{AAS\TeX}
\newcommand\latex{La\TeX}
\newcommand\kms{km$\,$s$^{-1}$}
\newcommand\Msol{M$_{\odot}$}
\newcommand\Lsol{L$_{\odot}$}
\newcommand\MHI{$M_mathrm{HI}$}
\newcommand{\hi}{H\,{\sc i}}

\received{\today}
\revised{\today}
\accepted{\today}

\submitjournal{ApJS}

\shortauthors{Zaritsky, et al.}{
\shorttitle{SMUDGes in Stripe 82}

\title{Systematically Measuring Ultra-Diffuse Galaxies (SMUDGes). II.  Expanded Survey Description and the Stripe 82 Catalog}
  
\correspondingauthor{Dennis Zaritsky}
\email{dennis.zaritsky@gmail.com}

\author[0000-0002-5177-727X]{Dennis Zaritsky}
\affiliation{Steward Observatory and Department of Astronomy, University of Arizona, 933 N. Cherry Ave., Tucson, AZ 85721, USA}

\author[0000-0001-7618-8212]{Richard Donnerstein}
\affiliation{Steward Observatory and Department of Astronomy, University of Arizona, 933 N. Cherry Ave., Tucson, AZ 85721, USA}

\author[0000-0001-8855-3635]{Ananthan Karunakaran}
\affiliation{Department of Physics, Engineering Physics and Astronomy Queen's University Kingston, ON K7L 3N6, Canada}

\author[0000-0002-5292-2782]{C. E. Barbosa} 
\affiliation{Universidade de S\~{a}o Paulo, Instituto de Astronomia, Geof\'isica e Ci\^encias Atmosf\'ericas, Departamento de Astronomia, Rua do Mat\~{a}o 1225, S\~{a}o Paulo, SP, 05508-090, Brazil}

\author[0000-0002-4928-4003]{Arjun Dey}
\affiliation{NOIRLab, 950 N. Cherry Ave., Tucson, AZ 85719, USA}

\author[0000-0002-3767-9681]{Jennifer Kadowaki}
\affiliation{Steward Observatory and Department of Astronomy, University of Arizona, 933 N. Cherry Ave., Tucson, AZ 85721, USA}

\author[0000-0002-0956-7949]{Kristine Spekkens}
\affiliation{Department of Physics, Engineering Physics and Astronomy Queen's University Kingston, ON K7L 3N6, Canada}
\affiliation{Department of Physics and Space Science Royal Military College of Canada P.O. Box 17000, Station Forces Kingston, ON K7K 7B4, Canada}

\author[0000-0002-0123-9246]{Huanian Zhang\begin{CJK*}{UTF8}{gkai}(张华年)\end{CJK*}}
\affiliation{Steward Observatory and Department of Astronomy, University of Arizona, 933 N. Cherry Ave., Tucson, AZ 85721, USA}

\begin{abstract}
We present 226 large ultra-diffuse galaxy (UDG) candidates ($r_e > 5.3$\arcsec, $\mu_{0,g} > 24$ mag arcsec$^{-2}$) in the SDSS Stripe 82 region recovered using our improved procedure developed in anticipation of processing the entire Legacy Surveys footprint. The advancements include less constrained structural parameter fitting, expanded wavelet filtering criteria, consideration of Galactic dust, estimates of parameter uncertainties and completeness based on simulated sources, and refinements of our automated candidate classification. 
We have a sensitivity $\sim$1 mag fainter in $\mu_{0,g}$  than the largest published catalog of this region.
Using our completeness-corrected sample, we find that (1) there is no significant decline in the number of UDG candidates as a function of $\mu_{0,g}$ to the limit of our survey ($\sim$ 26.5 mag arcsec$^{-2}$); (2) bluer candidates have smaller S\'ersic $n$; (3) most blue ($g-r < 0.45$ mag) candidates have $\mu_{0,g} \lesssim 25$ mag arcsec$^{-2}$ and will fade to populate the UDG red sequence we observe to $\sim 26.5$ mag arcsec$^{-2}$; (4)  any red UDGs that exist significantly below our $\mu_{0,g}$ sensitivity limit are not descendent from blue UDGs in our sample; and (5) candidates with lower $\mu_{0,g}$ tend to smaller $n$. We anticipate that the final SMUDGes sample will contain $\sim$ 30$\times$  as many candidates.
\end{abstract}    


\section{Introduction}
\label{sec:intro}

Recent findings of large, low surface brightness galaxies in large numbers
\citep[cf.][]{vdk15a,koda,mihos,munoz,roman17a,shi,vdb17,venhola,wittmann,greco,smudges,tanoglidis} have
renewed appreciation for these objects that first started puzzling us over 40 years ago \citep{disney,sandage,impey,schombert,sch,sprayberry,dalcanton,penny,conselice}. The largest among these galaxies are of a scale comparable, in certain ways, to that of our own galaxy \citep{vdk15a}. As such, they are challenging objects to understand in that they have an integrated star formation efficiency that is $\sim$ 1\% to 10\% that of our Galaxy. In the current vernacular, these large low surface brightness galaxies, commonly defined to have central $g-$band surface brightness fainter than
24 mag arcsec$^{-2}$ and effective radii greater than 1.5 kpc, are termed ultra-diffuse galaxies (UDGs). 

Although there is no evidence yet that they are not simply the long tail of the galaxy distribution toward both low central surface brightness and large physical size \citep{conselice,lim},  these are compelling objects for directed study. As shown in Figure 3 of \cite{vdk16}, they can lie in a region of parameter space that was previously empty \citep[this remains true even when the mass-to-light ratio of some of these are revised downward;][]{vdk19}. As such, they provide stress tests for models of galaxy scaling relations \citep{z08}, evolution \citep[e.g.,][]{amorisco16,cintio2017,carleton19,martin19,sales20}, and, possibly, dark matter behavior \citep{vdk18}.

With our observational program, we seek to compile as extensive a sample as currently possible of nearby, physically large UDGs across all environments to establish the characteristics of this population. We have coined our effort to be one of Systematically Measuring Ultra-Diffuse Galaxies or SMUDGes \citep[][hereafter Paper I]{smudges}.
In Paper I, where we first described the survey, we enumerated the reasons why the largest UDGs are extremely interesting test cases for dark matter studies, described our approach for recovering large numbers of such galaxies from existing data, and presented a catalog of UDGs in the region surrounding the Coma cluster as a demonstration case. Although the Coma cluster was also an area of focus for some of the earliest of the recent set of UDG studies \citep{vdk15a,koda}, UDGs are now known across all environments \citep[cf.][]{makarov,martinez,vdb17,roman17a,roman17b,greco,shi,wittmann,leisman,prole,tanoglidis} and here we also extend our survey to an area separate from the Coma cluster.

There are various shortcomings of our initial work that we need to address before proceeding to process the majority of the sky that is mostly not obscured by the Galaxy. First, in Paper I we did not provide a method for estimating the completeness of our survey as a function of position on the sky and UDG candidate properties. This omission severely limits the sample's value for statistical analyses of the UDG population and can obscure some key trends. Second, we did not present full estimates of parameter measurement uncertainties and biases. These are potentially quite severe for extremely low surface brightness objects \citep[e.g.,][]{vdb17, bennet}. Third, we did not address the deleterious effects of Galactic cirrus that become increasingly important as we study regions at lower Galactic latitudes than the Coma cluster region. While large swaths of dust are easily differentiated from UDG candidates, small, isolated regions of reflected light can produce false positive detections \citep{duc,roman19}.  Lastly, although we discussed some automated classification in the first paper, we depended heavily on visual classification. Such an approach is unmanageable when we extend our survey beyond regions that we are currently processing.

The core of the work remains unchanged. We still avail ourselves of the tremendous resource provided by the Dark Energy Spectroscopic Instrument project \cite[][and http://desi.lbl.gov]{schlegel,desi1,desi2} in the form of their Legacy Surveys imaging data \citep{dey}.
This paper describes these modifications in preparation for the upcoming presentation of the full catalog drawn from the entirety of the Legacy Surveys footprint. 
In \S\ref{sec:data} we describe the data. In \S\ref{sec:enhancements} we discuss in detail the enhancements we make to the data processing we described in Paper I. In \S\ref{sec:catalog} we present our catalog for the Stripe 82 area. All magnitudes are on the AB system \citep{oke1,oke2}. Finally, in \S\ref{sec:discussion} we discuss some preliminary inferences drawn from the catalog regarding UDG properties. 
For consistency with \cite{vdk15a}, we adopt an angular diameter distance of 98 Mpc to the Coma cluster, which implies a physical to angular scale of 0.475 kpc arcsec$^{-1}$ when referencing physical properties in the Coma cluster and vicinity. The corresponding luminosity distance to the Coma cluster is adopted to be 102.7 Mpc. For other calculations, we adopt WMAP9 cosmological parameters \citep{wmap9}.

\section{The Data}
\label{sec:data}

\begin{figure*}[ht]
\includegraphics[width=1.0\textwidth]{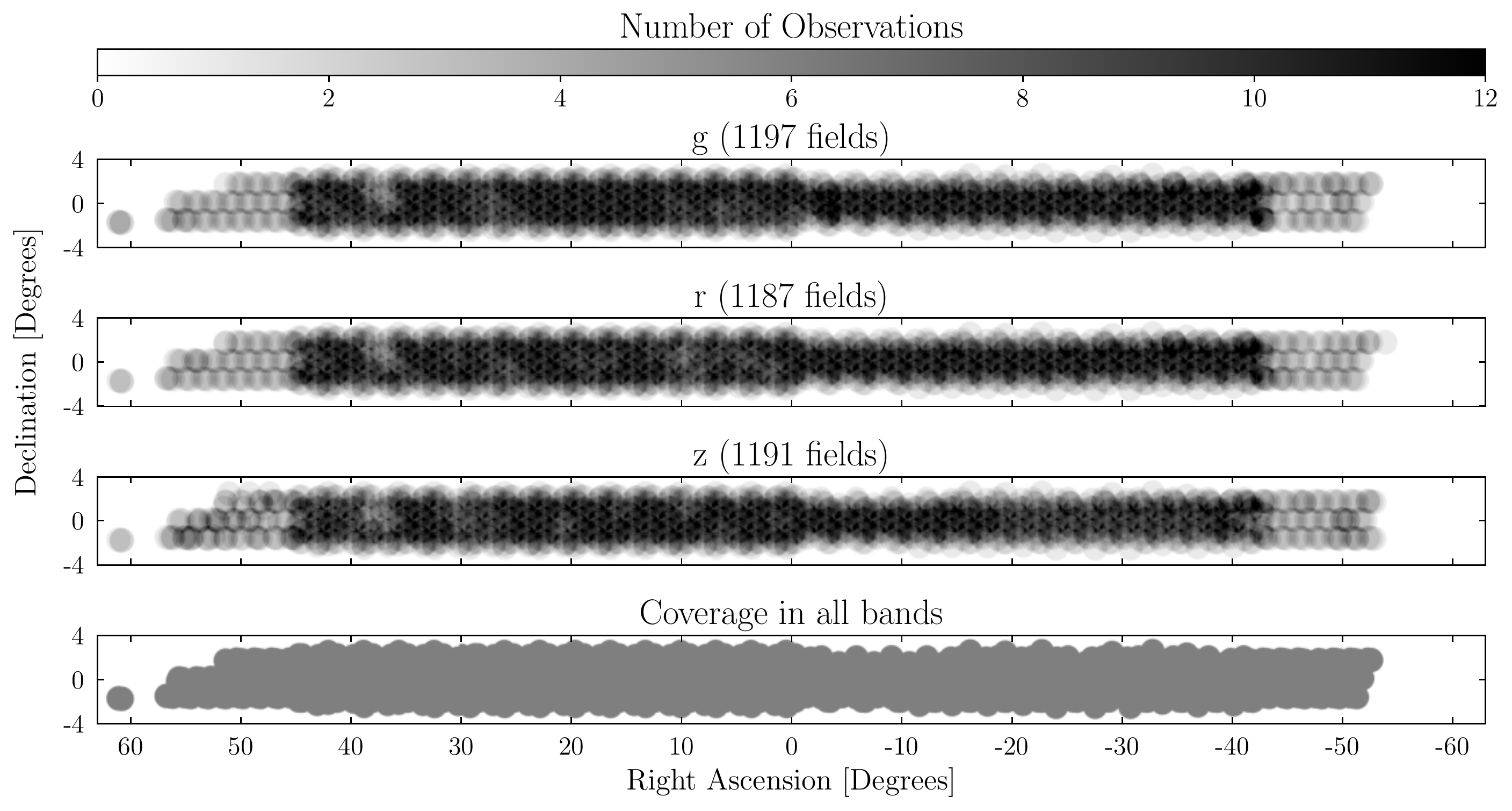}
\caption{Footprint of the Stripe 82 observations used in this study. In the top three panels, the shading denotes the observation density for each band as shown in the top color bar. The bottom panel shows regions for which we have coverage in all three filters.}
\label{fig:footprint}
\end{figure*}
Stripe 82 is a region encompassing $\sim$300 deg$^2$ along the Celestial Equator that has been repeatedly imaged by the Sloan Digital Sky Survey (SDSS) as described by \cite{annis} and \cite{Jiang}.  It extends from about 20h to 4h in right ascension and from $-$1.26$^\circ$ to $+1.26^\circ$ in declination.  We selected this region as the second test of our pipeline for several reasons. First, the available deeper imaging enables us to assess the limitations of our procedure on the more typical shallower data. Moreover, it has been extensively imaged by the Dark Energy Survey \citep{des} which is included within the DECaLS collection \citep{dey}.  Some objects were observed as many as 32 times, which allows us to estimate our detection limits independent of SDSS.   Second, other investigators are applying their techniques to search for low surface brightness objects in at least portions of this region \citep{fliri,roman18,greco,roman19,tanoglidis}, which will provide independent benchmarks. Third, multiwavelength complementary data exist, making it an excellent field region to explore \citep[e.g.,][]{hodge,xmm,nearIR,splus}. Lastly, it provides a physical contrast with our initial survey area that was centered on the Coma cluster and covered a similar sized area ($\sim 290$  deg$^2$; Paper I).

A detailed description of the Legacy Surveys is available in \cite{dey}. Briefly, the data consist of 
a 3-band  imaging survey, with $g$ = 24.7, $r$ = 23.9, and $z$ = 23.0 AB mag, 5-sigma point-source
limits, obtained using DECam at the CTIO 4-m (DECaLS), an upgraded MOSAIC camera at the KPNO 4-m (MzLS, Mayall $z$-band Legacy Survey), and the 90Prime camera at the Steward Observatory 2.3m telescope (BASS, Beijing-Arizona Sky Survey). These limits make it 
roughly 2 magnitudes deeper than SDSS, with better image quality as well.
The pipeline processed data, using the DECam Community Pipeline \citep{valdes}, are publicly available at the NSF's National Optical-Infrared Astronomy Research Lab (NOIRLab) science
archives\footnote{\href{https://astroarchive.noao.edu}{astroarchive.noao.edu}} and the Legacy Surveys' website\footnote{\href{https://www.legacysurvey.org/dr8/}{legacysurvey.org/dr8/}}. 

In this paper, we reanalyze  the CCD images associated with the 8th data release (DR8) of the Legacy Surveys. 
We focus on applying our own post-processing of the Legacy Surveys images to compile a catalog of low surface brightness ($\mu_{0,g} \ge 24$ mag arcsec$^{-2}$) galaxies of large angular extent (effective radii $r_e \gtrsim$ 5.3 \arcsec, corresponding to physical values $\ge$ 2.5 kpc at the distance of Coma) in the Stripe 82 field. As we went about this work, we reworked our pipeline to address the issues described above. 
The DECam field of view is 2.2$^\circ$.  To ensure that that we use all DECaLS observations covering the Stripe 82 footprint, we extend our analysis region by 1.2$^\circ$ on all sides and process all images with centers within these boundaries. This extension results in an area of $\sim$670 deg$^2$ with coverage in all three filters as shown in Figure \ref{fig:footprint}.
The Coma region that we discuss is described in detail in Paper I.

\section{Enhancements to our Post-Processing}
\label{sec:enhancements}

We perform our image processing and analyses using the Puma
cluster at the University of Arizona High Performance Computing center\footnote{\href{https://public.confluence.arizona.edu/display/UAHPC/Resources}{public.confluence.arizona.edu/display/UAHPC/Resources}}. A detailed description of our approach to building the UDG candidate catalog is given in Paper I. Here we focus on describing improvements and additions.  Our Stripe 82 footprint contains 3575 separate DECam exposures. Because of the amount of computer memory and temporary storage required for processing the entire footprint, the data are divided into four separate rectangular ``tiles" along the long axis of the stripe. These tiles are processed separately. To avoid difficulties with objects at the edges of the tiles and ensure that all of the images taken of each object are processed together in at least one tile, 
we overlap adjacent tiles by 1.2$^\circ$ in Right Ascension to account for the 2.2$^\circ$ DECam field of view.

A critical component of this work is the file survey-ccds-decam-dr8.kd.fits, which is included in the Legacy Surveys Data Release 8 (DR8), that contains information for each individual CCD image used in the data release and excludes those CCDs considered inadequate for further processing. We therefore limit our processing to the images included in that file. The file also contains magnitude zero points and image FWHMs generated for each CCD by the Legacy Surveys pipeline and we use these values when needed. The descriptions below apply to the processing of a single tile and the process is identical for all four tiles.  Unless otherwise stated, total numbers provided below indicate the total for all four tiles. Duplicate identifications of UDG candidates (hereafter, often referred to as candidates), from the overlap regions, are removed after all tiles are completely processed.

\subsection{Modifications to SMUDGes v1.0}

Our processing pipeline consists of the following major components. Steps that are entirely new to our procedure are written in bold while those that were modified from our previous work are in italics:

\medskip
\noindent
1) {\it identify and remove CCD artifacts;}

\smallskip
\noindent
2) {\it subtract high surface brightness objects using a model that includes background noise;}

\smallskip
\noindent
3) {\it detect candidates using wavelet filtering;}

\smallskip
\noindent
{\bf4) reproject images to minimize misalignments;}

\smallskip
\noindent
5) create a list of potential candidates by requiring coincident detections among different, overlapping, multi-band exposures;

\smallskip
\noindent
6) {\it obtain preliminary candidate parameters using a least-squares fit of an  n = 1 (exponential) S\'ersic model to each candidate on each individual image that produced a detection, and
reject detections that do not satisfy conservative size and brightness criteria;}

\smallskip
\noindent
7) obtain refined candidate parameters using GALFIT \citep{peng} and an $n = 1$ S\'ersic model on a stacked image for each candidate produced using all observations, again rejecting some using conservative criteria;

\smallskip
\noindent
{\bf 8) obtain final {\bf candidate} parameters using GALFIT, including now the effects of an estimated point-spread function (PSF) and fitting for the S\'ersic index;}

\smallskip
\noindent
{\bf 9) reject candidates from areas of sky where Galactic dust map values exceed predefined thresholds;} 

\smallskip
\noindent
10) {\it execute automated classification of all candidates and visually screen all those that remain viable}; and

\smallskip

\noindent
{\bf11)  add simulated UDGs to the images and reprocess the data to enable us to measure recovery completeness and estimate random errors and biases in our results.}

\smallskip\noindent
Below we briefly describe each step and, when pertinent, discuss why and how we have modified them from our previous implementation. Additions to the pipeline are described in detail.

\subsubsection{Image Preprocessing}
We process a total of 227,001 CCDs for Stripe 82.  Artifacts identified in the data quality mask provided by the DECam Community Pipeline are removed as described in Paper I. We then fit the extended wings of stars identified as saturated in these same masks with a Moffat profile and subtract the model from the image. In a change from Paper I, we no longer subtract wings of bright stars not identified as saturated because this adds significant processing time without noticeable benefit. Unsaturated stars are subtracted in a manner similar to other objects as described in the next section.

\subsubsection{Object Subtraction}
To reduce ``noise" before detecting candidates using wavelet filtering, we subtract foreground/background sources that have significantly higher surface brightness than our target UDGs.  Our previously imposed Source Extractor thresholds \citep{bertin} of 3$\sigma$ above sky occasionally also subtracted candidates in low-noise environments.  We correct this problem by now only subtracting objects that have a central surface brightness that is at least 2 mag arcsec$^{-2}$ brighter than a specified threshold in each band (24.0 for $g$, 23.6 for $r$, and 23.0 for $z$).  

\subsubsection{Wavelet Filtering}
We use wavelet transforms after object subtraction to isolate potential candidates of different angular scales with a tailored filter. Higher wavelet levels will preferentially accentuate objects of larger sizes.  In our previous work we selected candidates that were prominent at our level four (the smoothing kernel in this case has a FWHM $\sim$ 11.2$^{\prime\prime}$, see Paper I for details) because this choice highlighted the UDGs at the distance of the Coma region.  To detect potentially larger candidates, we now add both levels five and six to this process (each level increase is a factor of two in size). The filtering transforms the UDG candidates into highly statistically significant sources because they are not inherently faint sources, they just have their flux spread over many image pixels. The vast majority of the candidates turn out to have $m_g < 21$ mag. As such, we do not require the latest, most sophisticated image analysis tools to detect these objects in appropriately smoothed images (see Figure 4 in Paper 1 for a visual demonstration).

\subsubsection{Reprojection}
While not significant at the declination of Stripe 82 or the Coma cluster, image misalignment when stacking is an issue at the more extreme declinations included within the Legacy Survey. We minimize misalignment by using SWarp \citep{Swarp} to reproject all CCD images such that North is directly up before saving them for further processing. 

\subsubsection{UDG Candidate List Creation}
\label{sub:list}
After wavelet filtering, we have a total of 19,331,154 detections in Stripe 82, or an average of about 85 per CCD.  The vast majority of these will prove to be spurious, or otherwise poor  candidates.  Isolating, coadding and modeling  candidates is time-consuming and the following procedure  decreases the number of detections for which we need to perform each subsequent step. We  limit spurious detections by only retaining candidates with at least two coincident detections among different exposures, regardless of which wavelength filter was used in the detection.  Detections that meet this criterion must lie within 2\arcsec\ of the mean centroid of all the coincident candidate  detections. Each group of detections created in this manner is considered to be a unique candidate located at the mean centroid position. 

\subsubsection{Preliminary Parameter Screening}
\label{sub:prelim}
At this point we still have 3,528,371 distinct sources  containing a total of 11,811,322 detections that require further screening, the vast majority of which will not meet our UDG criteria. As described in Paper I, we limit the number of detections requiring time-consuming coaddition and GALFIT modeling by obtaining much faster, rough parameter estimates by fitting an exponential S\'ersic model ($n$=1) to each candidate on a CCD using the LEASTSQ function from the Python SciPy library \citep{jones}.  The fitting is done on a 201 $\times$ 201 pixel (54 $\times$ 54 arcsec) cutout image centered on the candidate location. 

As part of this process we mask objects not associated with the candidate detection to minimize their effects on the model. We create the mask from a segmentation map of a Gaussian-smoothed version of the cutout image using SEP \citep{sep}, an application based on Source Extractor \citep{bertin}. Because the cutout is centered on the candidate, we remove the central detection from the mask, thereby allowing it to be modeled. We found that our original detection threshold of twice the background noise occasionally missed very faint candidates and we now use a threshold of 1.5 times background noise.  This fitting provides only rough parameter estimates; therefore, 
we set requirements for candidates that are generous relative to our final criteria. For now we retain candidates with $r_e > 4$\arcsec\ and $\mu_0$ thresholds of 23.0, 22.0 and 21.5 mag arcsec$^{-2}$ for $g$, $r$, and $z$, respectively.

\subsubsection{Refined Parameter Screening}
\label{sub:GF1}
After preliminary screening we are left with 464,796 detections of 322,704 candidates, confirming our hypothesis that the vast majority of wavelet detections are spurious or do not meet our UDG criteria.  
Although we required at least two coincident detections for each candidate (\S\ref{sub:list}), following this initial screening some  candidates are left with only one viable detection. Again requiring confirmation of each candidate source, we reject any candidates with only a single remaining detection. 

The specific requirement of at least two coincident detections was based on our initial experience with our study of the Coma region, which occurred relatively early in the DECaLS survey when observations were limited. In Stripe 82, we are working in regions with more than 30 observations. In such cases, requiring only two detections is rather weak confirmation of the source. Therefore, we require that candidates pass our preliminary screening on at least 20\% of the available observations, or a minimum of two observations for those with less than ten observations. 

Again, the vast majority are rejected after applying these criteria and we are left with 20,058 
candidates that require image stacking and modeling.  These are performed on 201 $\times$ 201 pixel cutouts centered on the candidate's coordinates.  Details on preparing the cutouts for fitting are discussed in Paper I. For stacking the final image, we use all CCDs covering the coordinates rather than using only those that resulted in a detection of the candidate.  We now perform a slightly more sophisticated, but still compromised, fit using  GALFIT where the S\'ersic index remains fixed at $n = 1$ and the PSF is not incorporated into the model. These constraints are placed at this step for computational efficiency. 
In a change from Paper I, we start by using GALFIT to estimate morphological parameters ($b/a$, $\theta$, and $r_e$) using a cutout created by stacking all associated CCDs, regardless of band.  Photometric properties ($\mu_0$, magnitude and color) are then obtained from stacked images in each band with these morphological values held constant during modeling.  Because we will do a final pass using GALFIT with a floating S\'ersic index and incorporating the PSF, we again set our criteria generously at this stage, requiring only $r_e \ge 4\arcsec$, $b/a \ge 0.34$, and $\mu_{0,g} \ge$ 22.95 mag arcsec$^{-2}$ or $\mu_{0,z} >$ 21.95 mag arcsec$^{-2}$ if there is no available measurement of $\mu_{0,g}$.

\subsubsection{Final Parameter Determination}
\label{sub:Galfit_n}
After our initial GALFIT pass we are left with 6,625 candidates 
that require better modeling.  In Paper I we only used a model with a variable S\'ersic index for our confirmed candidates.  Now, we allow the S\'ersic index to vary for all candidates and include an estimated PSF in GALFIT. We create the estimated PSF using a Gaussian kernel with a standard deviation equivalent to the median FWHM obtained from DR8 for all CCDs included in the stack being processed.  Our UDG criteria of $\mu_{0,g} \ge 24$ mag arcsec$^{-2}$ (or $\mu_{0,z} \ge 23$ mag arcsec$^{-2}$ in a few cases where $g$ data are missing), $r_e \ge 5.3 \arcsec$ and $b/a \ge 0.37$ are applied to the results from this step. These are applied without consideration of uncertainties, either random or systematic. Systematic uncertainties are described in \S\ref{sec:simulation} and correcting for those will lead to the inclusion in our final catalog of a few sources that exceed these defined criteria.
Our final set of quantitative selection criteria is reprised in Table \ref{tab:criteria} and the candidate catalog is discussed and presented in \S\ref{sec:catalog}. Before proceeding to the catalog, however, we expand on some key new aspects of our approach.

\begin{deluxetable}{lr}
\caption{UDG Candidate Final Parametric Selection Criteria}
\tablewidth{\linewidth}
\label{tab:criteria}
\tablehead{
\colhead{Parameter}&
\colhead{Criterion}\\
}
\startdata
$\mu_{0,g}$&$\ge  24.0$ mag arcsec$^{-2}$\\
$r_e$&$\ge 5.3$\arcsec\\
$n$&$<$ 2\\
$b/a$&$\ge $ 0.37\\
$g-r$&$< 1$ mag\\
\enddata
\end{deluxetable}

\subsubsection{Coarse Screening of Spurious Sources Caused by Cirrus}

In contrast to the Coma region, significant portions of Stripe 82 are near the Galactic plane with resulting greater Galactic cirrus contamination.  Small regions of dust may reflect light and, without more information, it can be difficult to differentiate these low surface brightness enhancements from legitimate candidates. To address this challenge, we develop a screening process for probable cirrus contamination.  
Cirrus is more easily discerned using color images and larger regions than in the grayscale cutouts used for our modeling. We use color JPEG images retrieved from the archive 
and the Legacy Surveys Viewer\footnote{\href{https://www.legacysurvey.org/viewer}{https://www.legacysurvey.org/viewer}}.  Even with these additional images it can sometimes be  difficult to identify cirrus if it smoothly covers large areas. 
We visually inspected nearly 1000 Stripe 82 candidates meeting our UDG criteria for the presence or absence of dust and, or, a UDG. Although subjective, our dust classifications are adequate for our purposes because we
are aiming only to guide our application of a more quantitative approach. 
We visually classify 579 candidates as reflected light from cirrus, 26 as a UDG candidate superimposed on reflected light from cirrus, 156 as uncontaminated candidates, and 168 as neither (e.g., tidal tails, poorly resolved groups and clusters, scattered light, etc.).

\begin{figure*}[ht]
\includegraphics[width=1.0\textwidth]{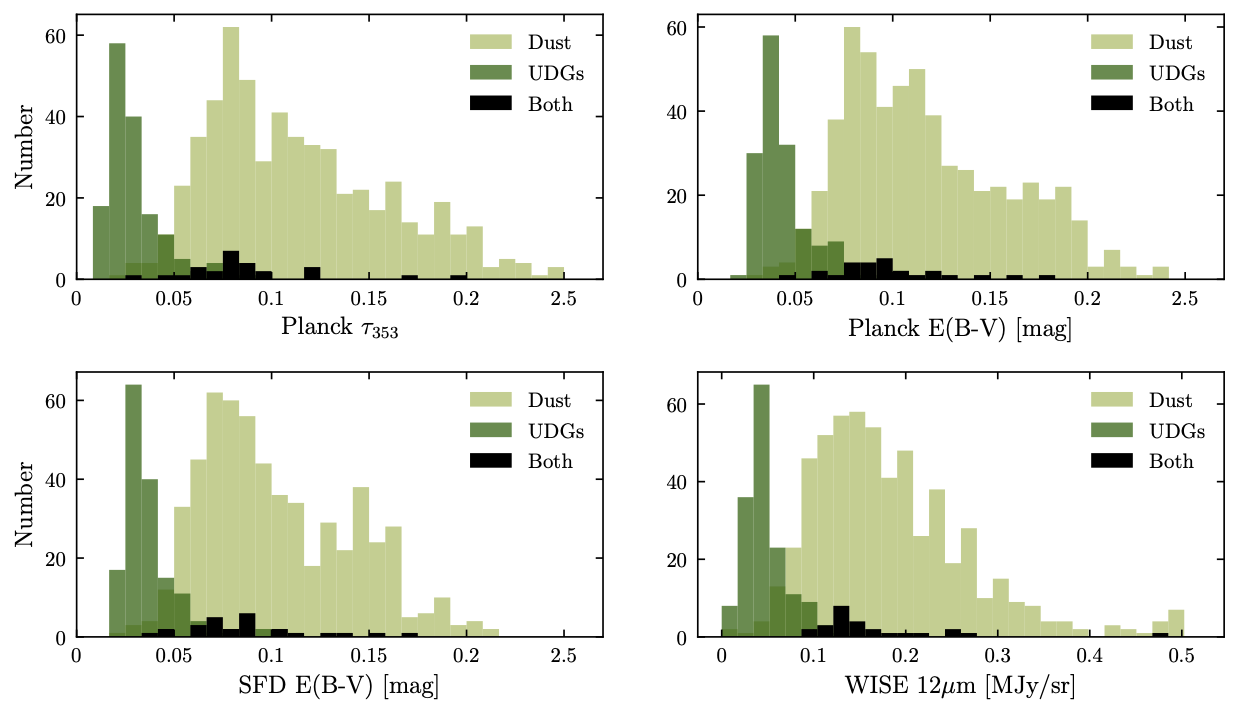}
\caption{Distribution of UDG candidate visual classifications (color coded) relative to four measurements of the column density of dust along the line of sight. For illustration purposes, any value of WISE 12 $\mu$m flux above 0.5 MJy/sr is set to 0.5. In all cases there is a relatively clean separation between candidates we visually classified as UDG candidates and sources attributed to reflected light from dust clouds.
}
\label{fig:dustUDG}
\end{figure*}

We consider four existing Galactic dust maps for our quantitative screening procedure. Three are provided by dustmaps.py \citep{green} and include $\tau_{353}$ (optical depth at 353 GHz), $E(B - V)$ from the Planck Collaboration \citep{planck}, and the Schlegel, Finkbeiner, Davis (SFD) dust map \citep{SFD}.  The fourth map is derived from WISE 12 $\mu$m observations by \cite{meisner}.  We extract single point values from each dust map located at the coordinates of each candidate. For the WISE values, we set any map value that is $<$0 to 0. We refer to these measurements as dust proxies.

Results comparing the various map values to our classifications are shown in Figure \ref{fig:dustUDG}, where we plot the distributions of the various dust proxy values at the position of candidates we identified as cirrus, UDG  candidate, and candidate + cirrus.  All four plots are quite similar, suggesting the choice of dust proxy does not play an integral part in the results.  As expected, the peak of the proxy value distributions associated with candidates unambiguously classified as UDG candidates lie at low values. Conversely, the distribution of candidates that we classified as arising due to dust have higher dust proxy values. Fortunately, the two regimes appear quite distinguishable with only modest overlap. 

Using the above information, we choose to select values from the $\tau_{353}$ and WISE 12 $\mu$m dust maps to set our thresholds.  Although all four dust proxies appear to work similarly well in distinguishing UDGs from reflected light sources, we opt to use the two measures that do not depend on extinction curves and $R$ values,  i.e., the WISE 12$\mu$m and $\tau_{353}$ estimates. In Figure \ref{fig:TauWise} we show where each candidate confirmed as a UDG candidate, dust, or candidate superimposed on dust falls on the WISE 12 $\mu$m-$\tau_{353}$ plane.  Although there is some overlap, candidates and dust primarily fall into two distinct groups.  Selecting a threshold is a compromise between rejecting legitimate candidates and allowing the possibility of isolated, small dust patches to be misclassified as potential UDGs. We set thresholds of $\tau_{353} = 0.05$ and WISE 12 $\mu\mathrm{m} = 0.1\,$MJy/sr as shown by the dotted lines in Figure \ref{fig:TauWise} and reject candidates with dust proxy values  exceeding either threshold. Even so, as is evident from Figures \ref{fig:dustUDG} and \ref{fig:TauWise}, these criteria fail to exclude all dust contamination.

We did not screen for dust contamination in Paper I.  We do that now for the 275 catalog entries presented there and find that 5 (1.8\%) exceed our thresholds.  We reviewed each of these and found one, SMDG1257423+211254, that appears to be only cirrus. 
The other four are SMDG1231377+203617, which barely failed our threshold ($\tau_{353}$ = 0.05003; WISE 12 $\mu$m = 0.04608) and which we feel is a faint UDG, as well as SMDG1301004+210356, SMDG1302280+204900, and  SMDG1304250+210738 which appear to be UDGs superimposed on dust.

Examining the WISE 12 $\mu$m dust map over the entire DECaLS footprint in the Legacy Survey Sky Viewer\footnote{https://www.legacysurvey.org/viewer} suggests that Stripe 82 has above average cirrus contamination while the Coma region has less than average. Using dust proxy map values at the locations of our randomly placed simulated sources as described in \S\ref{sec:simulation}, we find that 43.4\% of the Stripe 82 region and 1.7\% of Coma exceed our dust thresholds (Figure \ref{fig:DustFP}), confirming the large variation within the DECaLS footprint.  Our completeness estimates discussed in that Section will account for these losses due to cirrus, but it is evident that dust is likely to be a dominant source of large-scale completeness variations. 

\begin{figure}[ht]
\begin{center}
\includegraphics[width=0.45\textwidth]{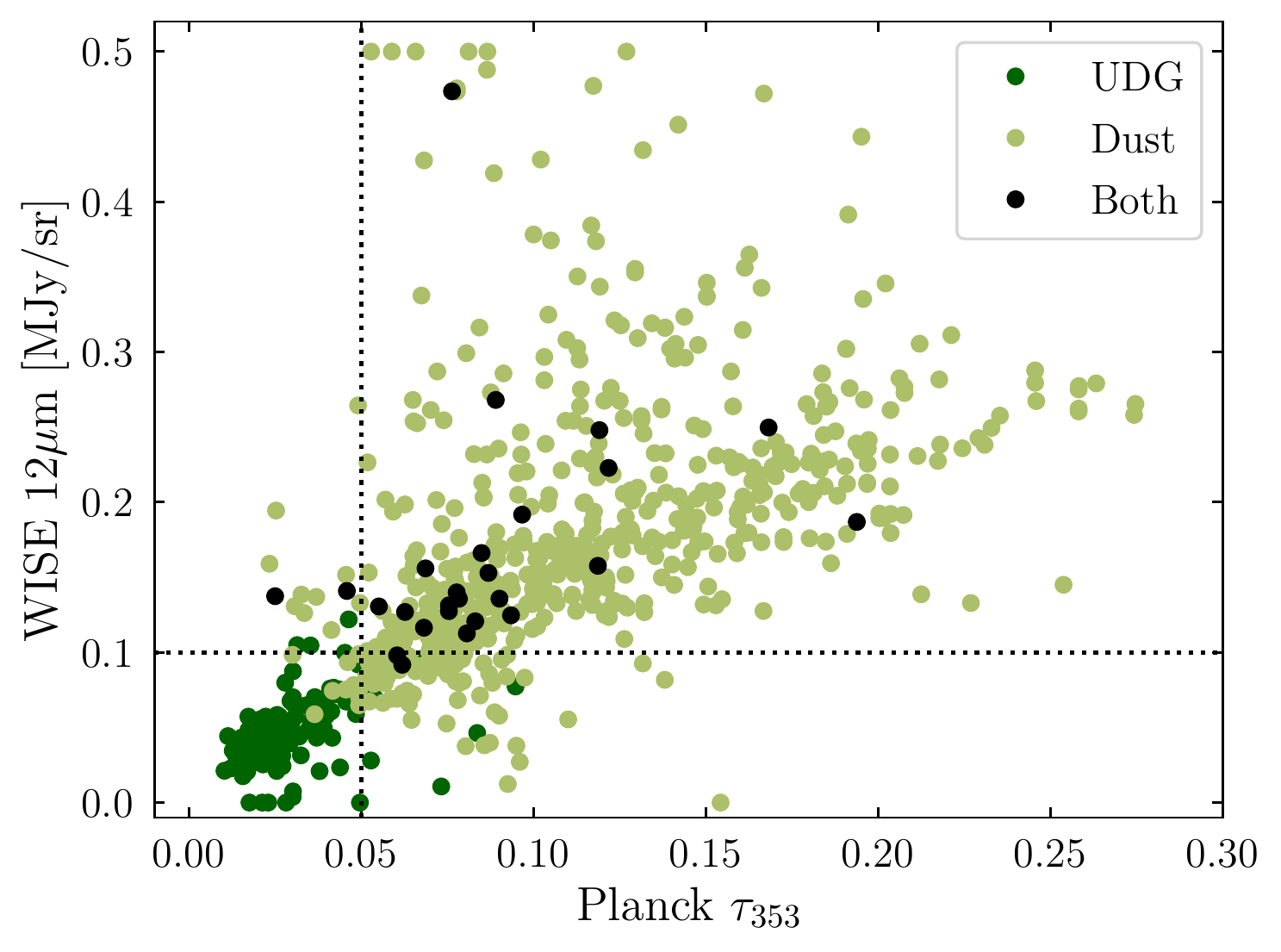}
\end{center}
\vskip -.5cm
\caption{WISE 12 $\mu$m 
flux vs. Planck-derived optical depth ($\tau_{353}$) at the locations of
UDG candidates.
Our independent visual classification of
the candidates 
is shown in the color coding. 
Dotted lines represent the thresholds we have set using this comparison for rejecting candidates on the basis of the WISE and Planck measurements. Only candidates in the lower left quadrant are retained.}
\label{fig:TauWise}
\end{figure}

\begin{figure}[ht]
\begin{center}
\includegraphics[width=0.47\textwidth]{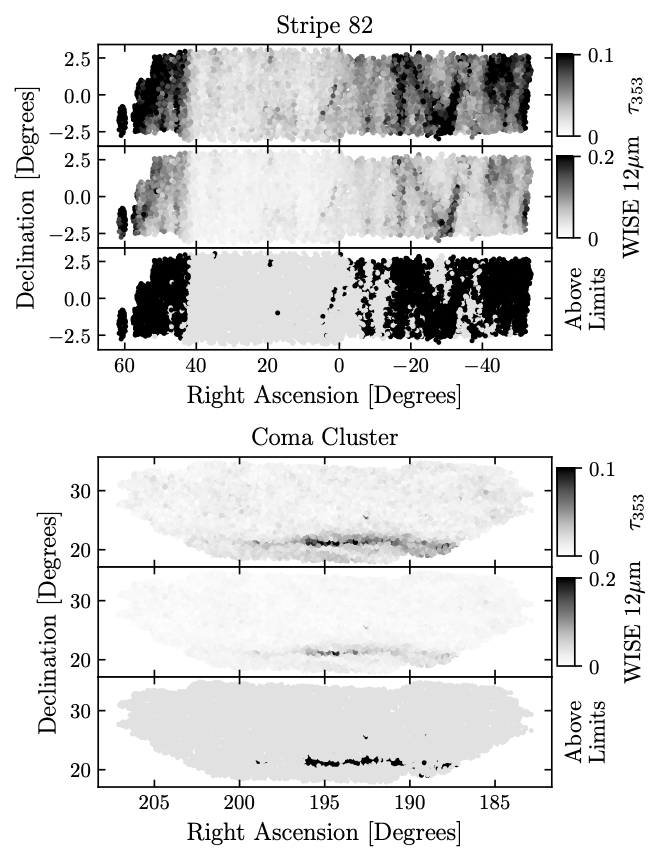}
\end{center}
\vskip -.5cm
\caption{Cirrus contamination in our Stripe 82 (top three panels) and Coma cluster (bottom three panels) footprints. The top panel in each set shows the distribution of $\tau_{353}$ while the middle panels show it for WISE 12 $\mu$m.  In the bottom panels in each set, the regions in black exceed our dust proxy thresholds (43.4\% of the area for Stripe 82 footprint and 1.7\% for Coma) and illustrate the large variation in dust contamination found within the DECaLS footprint. Due to plotting resolution areas that appear completely black are not necessarily uniformly so.}
\label{fig:DustFP}
\end{figure}
\subsubsection{Improved Automated Classification}
\label{sub:ML}

The argument of the relative merits of visual vs. automated classification becomes moot when dealing with the expected volume of candidates in the full version of SMUDGes. At the current scale, we can still contemplate visual screening, but this cannot scale further. In addition, as discussed in \S \ref{sec:simulation}, we are  increasing the size of our sample further by injecting artificial sources.  As such, the classification problem is multiplied many-fold. Automated classification becomes an imperative.  

To maximize the data set available for training, we combine the Coma candidates from Paper I with those from our current processing of Stripe 82.  Because of processing changes, we reanalyze our Coma observations using our current pipeline version and revised criteria.  This results in a total of 3073 candidates (1568 from Stripe 82 and  1505 from Coma) that are successfully modeled by GALFIT using a variable S\'ersic index and a PSF correction (\S\ref{sub:Galfit_n}). 
To further increase the number of UDGs in our classification training set and allow for simulated sources that may not meet our final thresholds, we set weaker criteria than those we apply in defining our final catalog. We will use all candidates that passed our initial screening using GALFIT fitting with a fixed S\'ersic index. 
Because many of the systems we are now classifying will not meet our final UDG criteria for reasons other than visual appearance, we will visually classify them as {\sl potential} UDG candidates and non-UDGs.  These are visually reviewed by three of the authors (DZ, RD, AK) and assigned to a class using the majority of our votes.  

During visual inspection we found a few objects that were structurally similar to other candidates but significantly redder than the Coma cluster red sequence. Because we do not expect to find high redshift low surface brightness galaxies nor do we expect UDGs to be highly reddened, we suspect that these are high redshift, high surface brightness galaxies masquerading as UDGs or possibly something more exotic such as emission line nebulae \citep[e.g.,][]{steidel}. Because it might be difficult for our machine learning algorithm to reject these with the current small sample of such objects, we define a $g - r$ color threshold instead.  As shown in Figure \ref{fig:red_sequence}, the vast majority of our confirmed UDGs within the Coma environs fall at or below the color of the bright tip of the Coma cluster red sequence. Using this result as a guide, we set a relaxed color criterion of $g-r < 1.0$ mag to exclude a small number of likely high redshift interlopers (Figure \ref{fig:color_hist}).  We do not exclude those where GALFIT failed to provide an estimate in either the $g$ or $r$ band.

Our visual classification also revealed a few cases where GALFIT incorporated a superimposed or abutting object in its model due to incomplete masking.  This results in parameter estimates with high S\'ersic indices and effective radii that are much larger than those suggested by visual inspection. We show and fit the distribution of measured $n$ values with a Gaussian ($\mu =0.89$ and $\sigma = 0.27$) in Figure \ref{fig:Sersic_n}. 
Based on this distribution, and seeking to remove the most egregious cases where the fit is affected by projected neighbors, we reject candidates that are best fit by models with $n \ge 2.0$.
With the new criteria just described ($g-r < 1.0$ mag and $n < 2.0$), we have a sample of 2665 candidates where 1539 are visually classified as potential UDG candidates and 1126 as non-UDGs.

Details on our approach for computer classification are described in the appendix of Paper I and here we only summarize and discuss slight differences in the approach and the results.  Because we want to evaluate different algorithms and hyperparameters, we set aside 20\% (533) of the candidates (306 potential UDG candidates and 227 non-UDGs) to be used as a test set after all hyperparameters are finalized.  The remaining candidates are divided into four folds to be used for cross-validation during evaluation.  In Paper I we used the coadded cutout images created by our pipeline.  Because one of the parameters we want to evaluate is image size, we now extract images centered on the candidates from the Legacy Surveys, which allows the needed flexibility.  We evaluate images ranging from 150 pixels ($\sim$39\arcsec) on a side to 300 pixels ($\sim$79\arcsec). We also test the relative effects of resolution vs. signal-to-noise ratio by rebinning the images by factors of 2 or 3. Because our data set is reasonably balanced between potential UDG candidates and non-UDGs, we augment the training set uniformly for each candidate by adding 4 images that are randomly flipped and/or rotated.  The FITS cutouts provided by the Legacy Survey are in units of nanomaggies and sky-subtracted (some pixel values may be $<$ 0) and we normalize these by clipping at $-$0.01 and 0.1 nanomaggies. 

We evaluate five different convolutional networks: the TensorFlow Keras versions of DenseNet121, DenseNet201 \citep{densenet}, EfficientNetB0, EfficientNetB1, and EfficientNetB2 \citep{efficientnet}.  As in Paper I, final layers are replaced with dense layers having a sigmoid activation as a single output. However, we no longer use pre-trained weights during initialization.  We found the best cross-validation results using EfficientNetB1 with a 224 $\times$ 224 pixel ( $\sim 59 \arcsec \times 59 \arcsec$) image without any smoothing and we choose this network configuration for classifying our final test set. Pertinent hyperparameters for this network are 1) dropout fraction of 0.3, 2) initial learning rate of 0.0001 which is decreased by multiplying by 0.3 every 20 epochs, 3) optimizer using Adam \citep{king} with beta\_1 = 0.9, beta\_2 = 0.999, epsilon = $10^{-7}$, and decay = 0.0, and  4) training length of 60 epochs with a batch size of 16 samples.  We do not use early-stopping at this time and the model shows signs of overfitting at the end of 60 epochs with the vast majority of probabilities approaching 0 or 1. However, the best accuracies during training are very noisy and only a fraction of a percent better than those at the end.  We may change our approach when we acquire a larger training set as we process the remainder of the Legacy Surveys.  Because we want to minimize false entries in our catalog, we set a probability threshold of 0.99 for accepting a candidate as a potential UDG. Using this threshold for our test set, we obtain an accuracy of 96.2\% (513/533) with  8 false positives (specificity of 96.5\%) and 12 false negative assignments (sensitivity of 96.1\%).

Images of all incorrect classifications are shown in Figure \ref{fig:misses}.  Classification errors tend to fall into a few main categories with potential UDG candidates confused with tidal material or outer spiral arms (c, d, g, l, t), distant faint clusters or apparently empty sky (e, i, j, m, p, q, r),  distant galaxies (a, f, s), and extended glow from a nearby object (n, o).  On occasion, the host galaxy for tidal material is not visible on the image supplied to the network, making automated classification very difficult.  In six cases (c, d, f, n, p, q) there were disagreements among us with the label assigned to the majority vote.  Panel (b) clearly shows cirrus contamination when a larger area is viewed, although we can not definitely rule out the possibility of a superimposed UDG.  This particular candidate barely missed our dust thresholds for rejection with $\tau_{353}$ = 0.048 and WISE 12 $\mu$m = 0.099 MJy/sr.  There were no similarly contaminated images in the training set.  The candidate in panel (h) was mislabeled during data entry and was correctly identified as a potential UDG by the network. It is included in our catalog.  The remaining false negative prediction in panel (k) barely missed our inclusion threshold with a probability of 0.9897. 

\begin{figure}
\begin{center}
\includegraphics[width=0.45\textwidth]{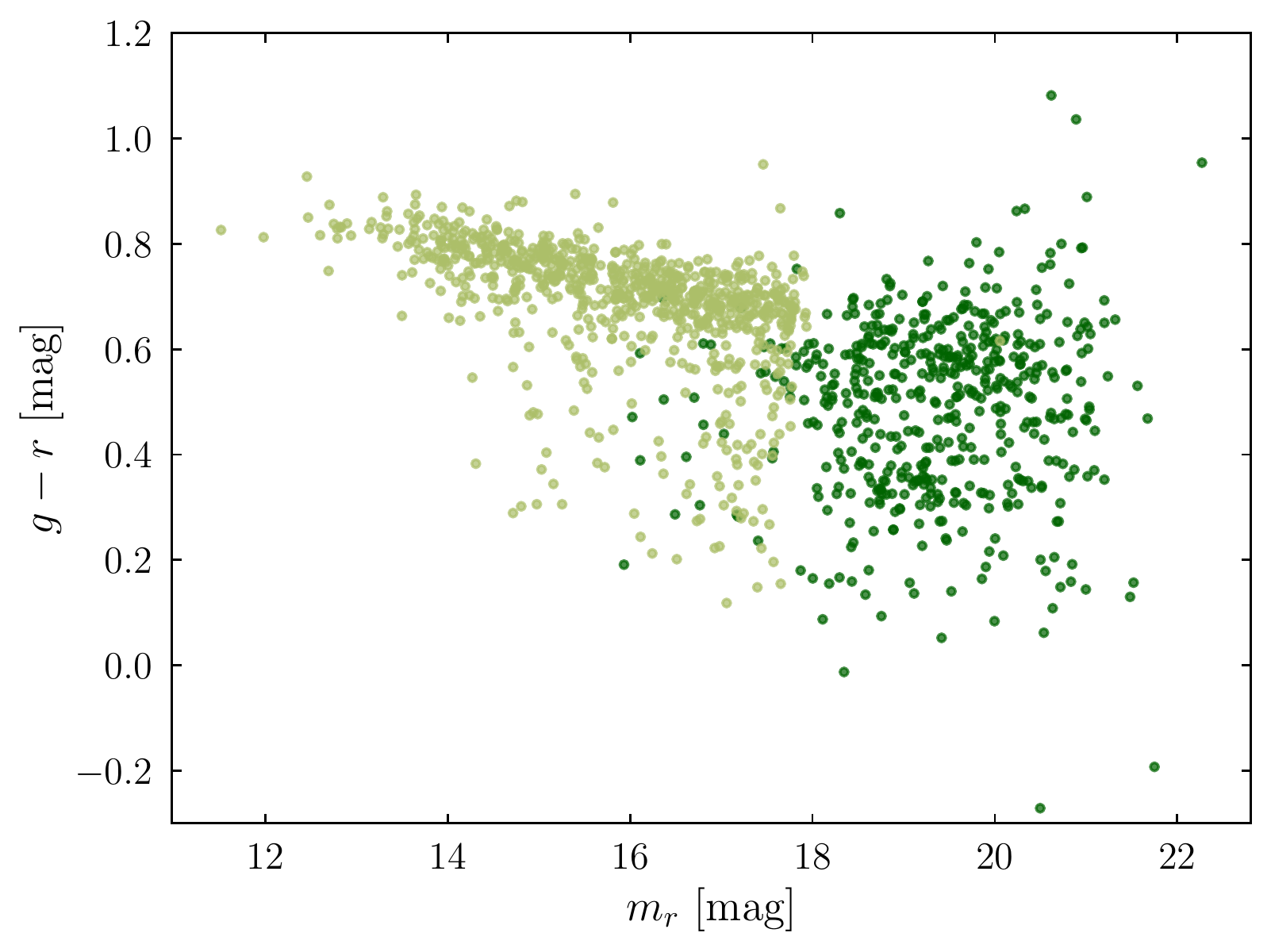}
\end{center}
\caption{Extinction-corrected $g - r$ color vs. apparent magnitude $m_r$ for SDSS galaxies within 2$^\circ$ of the Coma cluster that have spectroscopic redshifts between 0.018 and 0.028 (light green) and all of our UDG candidates within the Coma environs (dark green). }
\label{fig:red_sequence}
\end{figure}

\begin{figure}[ht]
\begin{center}
\includegraphics[width=0.45\textwidth]{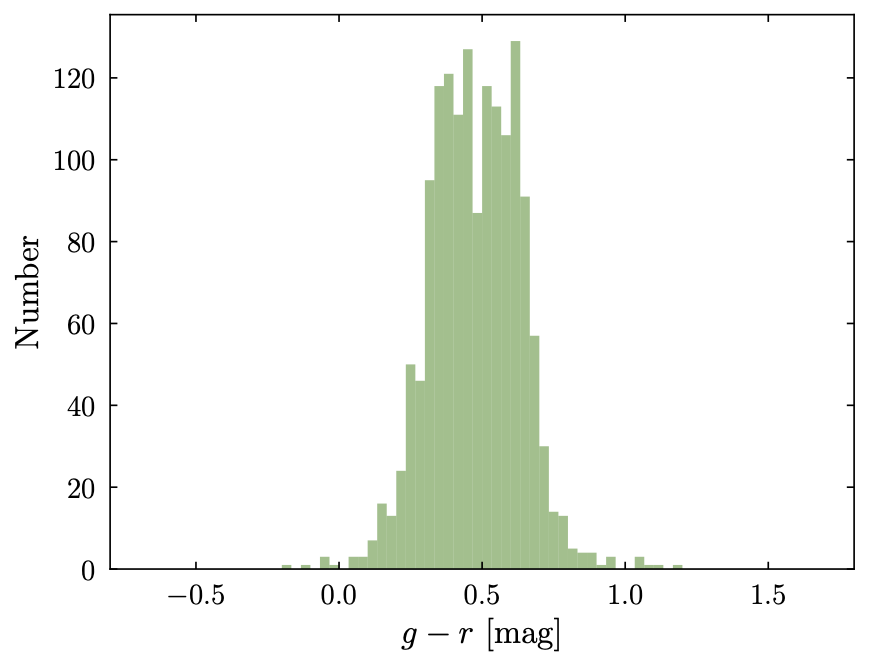}
\end{center}
\caption{Distribution of extinction-corrected $g - r$ colors of UDG candidates. The break between the red and blue populations can be seen as the dip in the color distribution at $g-r \sim 0.5$. The final catalog has additional criteria imposed that marginally affect this distribution.}
\label{fig:color_hist}
\end{figure}

\begin{figure}[ht]
\begin{center}
\includegraphics[width=0.45\textwidth]{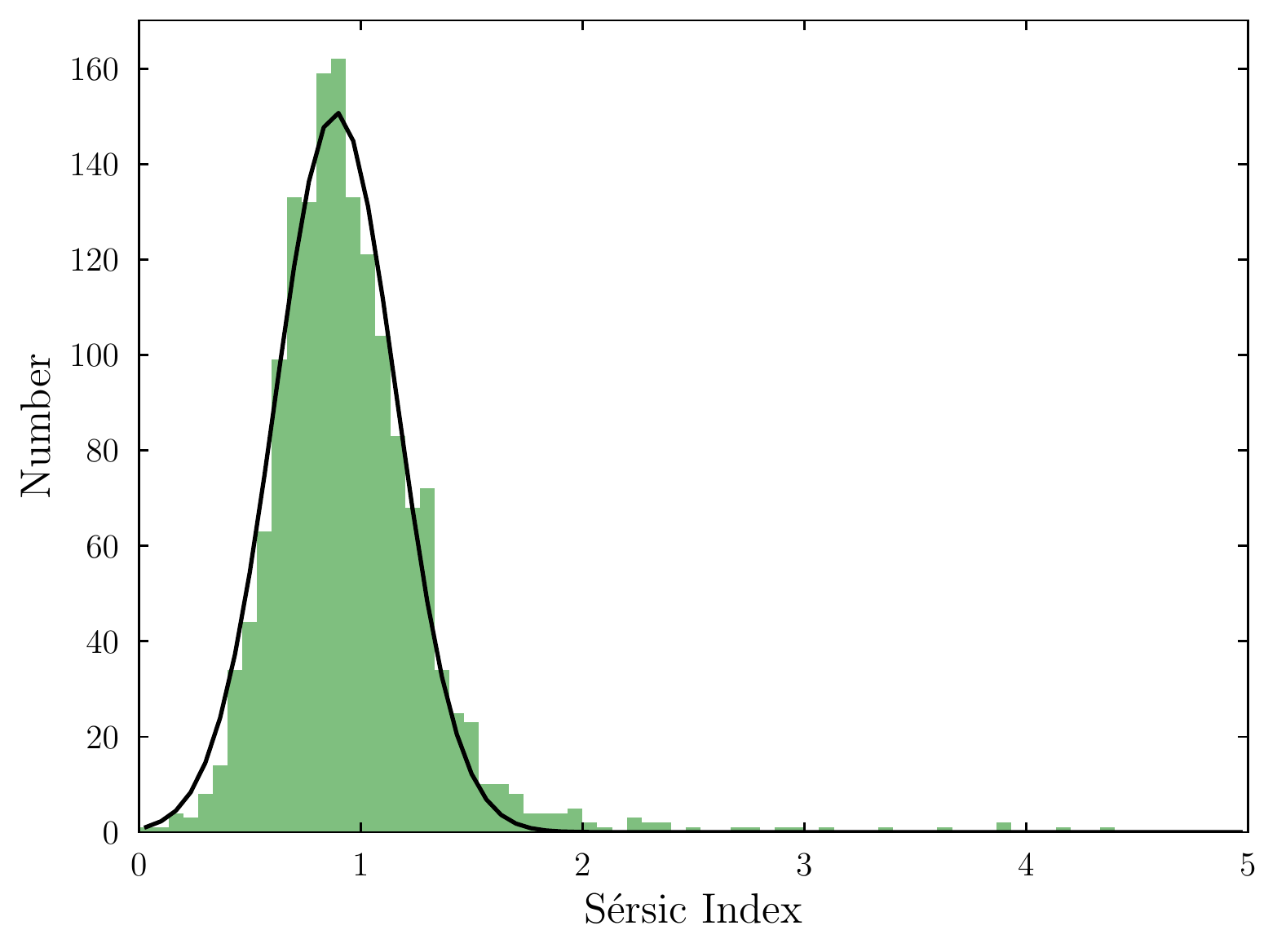}
\end{center}
\caption{S\'ersic n index distribution of UDG candidates. The solid line represents the least squares Gaussian fit for $0.5 \le n \le 1.5$, which has $\mu = 0.89$ and $\sigma = 0.27$ and is used to guide our definition of the range of acceptable $n$ values (see text). The final catalog has additional criteria imposed that marginally affect this distribution.}
\label{fig:Sersic_n}
\end{figure}

\begin{figure*}[ht]
\includegraphics[width=1.0\textwidth]{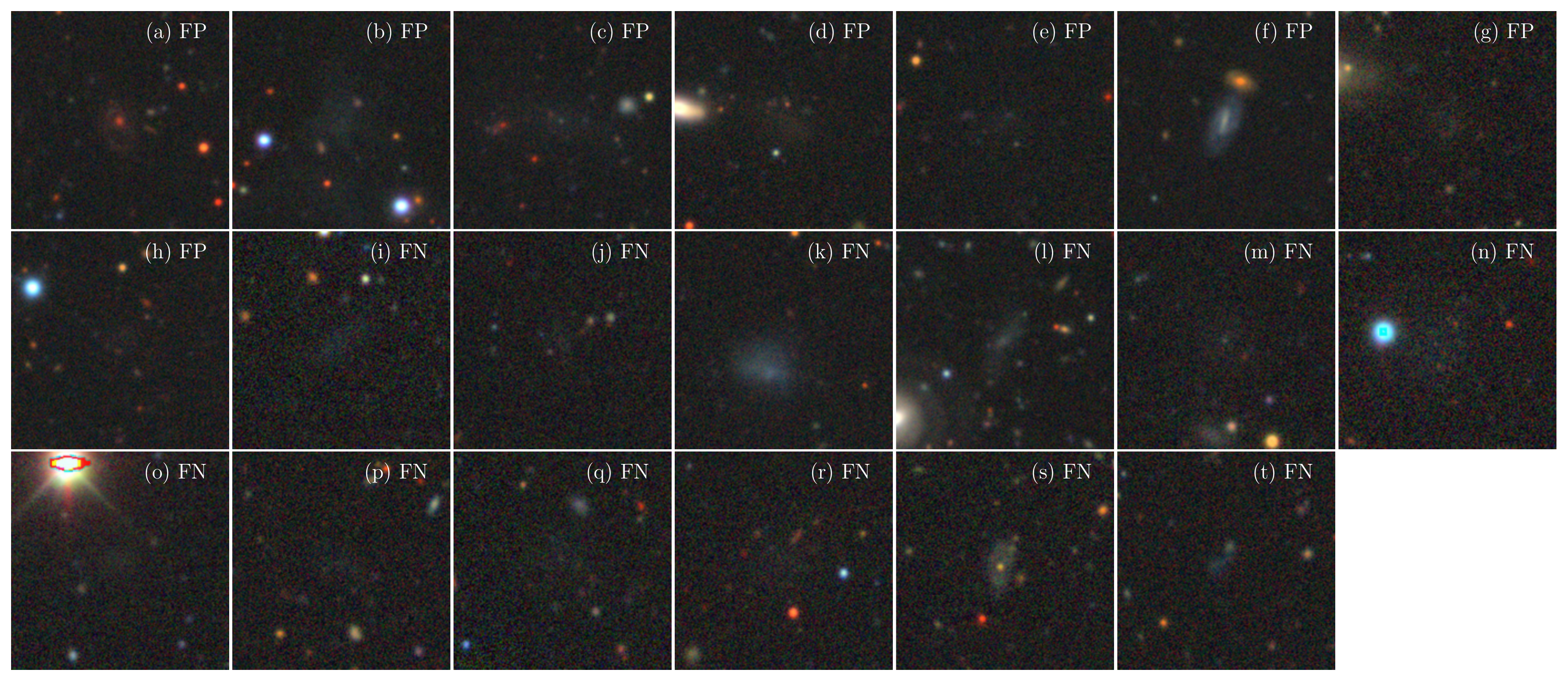}
\caption{Classifications where predictions produced by an automated classifier differ from our visual classification (FN = false negative; FP = false positive relative to our visual classification). RA, DEC, and name (if included in our catalog) are provided for each panel. See the text for details. (a) 357.76977, $-$0.18146. (b) 320.88322, 1.26719. (c) 20.30798, 1.38641. (d) 9.26339, $-$2.13205. (e) 358.39504, $-$0.79389. (f) 31.10827, 2.25552. (g) 195.04037, 20.44296. (h) 28.77117, $-$0.14588. (i) 194.12349, 28.44472; SMDG1256296+282641.  (j) 192.01841, 17.96964; SMDG1248044+175811.  (k) 324.99352, $-$2.41781; SMDG2139584-022504.  (l) 12.4503, 0.96312; SMDG0049481+005747.  (m) 194.95842, 28.1958; SMDG1259500+281145.  (n) 195.67625, 18.86472; SMDG1302423+185153.  (o) 192.835, 27.22306; SMDG1251204+271323.  (p) 183.34124, 29.56713; SMDG1213219+293402.  (q) 192.90708, 27.345; SMDG1251377+272042.  (r) 195.81109, 27.61941; SMDG1303147+273710.  (s) 193.21574, 22.30942; SMDG1252518+221834.  (t) 184.22557, 29.90266; SMDG1216541+295410. Color cutouts obtained using resources provided by the Legacy Surveys (\textit{Legacy Surveys Image Viewer/ D. Lang (Perimeter Institute))}}
\label{fig:misses}
\end{figure*}

\subsubsection{Simulated UDGs}
\label{sec:simulation}

Simulated sources were not added to the images in Paper I.  In this study, we estimate parameter systematic and random uncertainties, and recovery completeness by planting simulated UDGs at random locations throughout the observed region. These are modeled using S\'ersic profiles with random structural and photometric properties. To obtain an adequate number of simulated sources for robust analysis, we do the implanting and analysis separately from our science processing.  This approach allows us to use a higher density of simulated sources with no risk that the simulated sources affect the detection and measurement of real sources.

Although our approaches to estimating completeness and uncertainties differ in several aspects, they also have much in common and we discuss those commonalities first.  In both cases, image processing and classification are identical to that applied to our science images. To provide the most general results and maximize the number of available simulated sources, we combine results from both the Coma and Stripe 82 data.  

We randomly draw  S\'ersic indices ($n$), axis ratios ($b/a$), position angles ($\theta$), effective radii ($r_e$), central surface brightnesses in $g$ ($\mu_{0,g}$), and colors 
($g-r$, $r-z$) uniformly within the parameter ranges given in Table \ref{tab:uncertain_params}.  
Because errors may scatter objects into our selection space (as well as out), the parameter space explored extends beyond our UDG criteria when possible. The effective radius limits match the angular sizes of objects ranging from about 1.7 to 9.5 kpc at the distance of Coma. Recall that our minimum angular size criterion corresponds to 2.5 kpc at the distance of Coma, rather than the more standard 1.5 kpc UDG criterion. The  
color ranges represent the 3$\sigma$ limits of the distributions of our actual UDGs.  We randomly place simulated sources with unique identification numbers at an average density of 2000 per deg$^{-2}$ (about 100 per CCD) and require a minimum of 40\arcsec\ separation.  The source  locations are limited to the actual footprints of the surveys, including CCD gaps, and we require that any simulated source fall on at least 2 CCDs for inclusion. This last limitation is necessary because some observations at the edges of the footprint may not have any overlap and our standard pipeline would automatically reject any candidates in such regions.  CCDs are initially processed individually and any simulated source whose center falls within the boundaries of a CCD is placed on that CCD before any further processing.  As noted in \S\ref{sub:Galfit_n} describing our revised treatment, we now obtain our final GALFIT results for each candidate using a floating S\'ersic index and an estimate of the PSF.  Therefore, before placing a simulated source  on a CCD we convolve it with a Gaussian estimate of the PSF, using the point source FWHM provided by DR8 for each CCD. A detection by our pipeline that lies within 4\arcsec\ of a placed source is assigned the identification number of that source.  This separation tolerance allows some error in the location estimate, while still staying within our threshold for $r_e$. About 2.5\% of the detections are duplicates associated with the same simulated source and we simply select the first match and delete the others.

Finding that the surviving number of faint simulated sources was inadequate, we augmented the initial runs with two more where we created sources drawn from
a normal distribution of $\mu_{0,g}$ centered at 26.4 mag arcsec$^{-2}$ with a sigma of 1.3 mag arcsec$^{-2}$.  With these criteria a total of 3,700,777 simulated sources are generated, 2,772,724 fall completely on at least 2 CCDs, and 741,642 survive the pipeline through GALFIT modeling.

After applying our various criteria, including rejecting areas where dust contamination is suspected, we have a total of 440,015 surviving candidates before automated classification. As with our standard pipeline, we use 224 $\times$ 224 pixel cutouts obtained from the Legacy Surveys for this step.  
When we place a  simulated source on a cutout, we convolve the model with a Gaussian PSF created with a FWHM that is the mean FWHM of all of the CCDs contributing to the stacked image in each band.  Because of the high density of simulated sources, there may be cases where a real source is incorrectly associated with a simulated one, leading to errors in our estimates. We mitigate this problem by making two passes through our automated classification network (see \S \ref{sub:ML}).  In the first pass we use a cutout that does not include the simulated source associated with the detection and in the second, one that does. If we find a source classified as an UDG candidate on the first pass, then the association with the simulated source is rejected because we had not yet placed the simulated source at that location. In fact, in only 1,124 cases out of 440,015 (0.25\%) was there a source classified as a candidate that was sufficiently coincident on the location of the simulated source to cause potential confusion. We conclude that overlaps are an insignificant problem for our recovered measurements of simulated sources.  A total of 316,621 simulated candidates survive the automated classification and form the data set used for further evaluation.

In all cases, our models focus on what we consider the key parameters appropriate for exploring the variable under consideration.  These are  $\mu_{0,X}$ or $m_X$, $r_e$, $n$, and $b/a$, where $X$ represents each of the different filter bands. Specifically, we consider the structural components ($r_e$, $n$, and $b/a$) to be color-independent. We assess completeness with regard to $\mu_{0,g}$ and have confirmed that we see no effective change in completeness with color over the UDG color range.

To provide the measurements of completeness and uncertainties that will guide our modeling, we evaluate those quantities for simulated sources within limited range parameter windows that are repeatedly placed randomly throughout the parameter space.
There is a tension between setting smaller windows, which will contain more homogeneous systems and have higher resolution in tracing how quantities vary with parameter changes, and larger ones, which will have more robust statistics. To address this tension, 
we use a fixed window size in general but let it decrease to as small as 1/3 of its base size near the limits of a parameter range. We refer to each set of windows that describes the selection of sources in the 4-D parameter space as a bin.
In theory there is no limit to the number of bins that can be created, but the size of the data set must be large enough to allow adequate modeling across the full parameter space. The specifics of this basic implementation when we evaluate uncertainties and completeness are described in more detail in their respective sections.

In practice we have two distinct data sets that we will be discussing: 1) the parameters describing the unprocessed sample of simulated sources, and 2) the parameters recovered by our pipeline describing those same simulated sources. For clarity we refer to these as our input and output simulated samples, respectively.  

We fit the resulting multi-parameter data from our random binning technique using polynomial models created with the PolynomialFeatures function from the Python Scikit-learn library \citep{sklearn} and a four layer neural network implemented with Keras \citep{keras}.  To confirm that we are neither significantly under or overfitting the data, we use 80\% of the bin data set for training and set aside the remainder for testing and optimizing the neural network. Because the parameter ranges of our simulations (Table \ref{tab:uncertain_params}) exceed those of our UDGs, we have options for the selection of the limits used for modeling.  Broader limits will encompass a larger parameter space but may adversely affect the model.  We explore these limits separately for completeness and uncertainties.
A disadvantage of polynomial modeling, especially when using polynomials of high degree, is that the fit may extrapolate very poorly for data points lying outside of the fitted range.  We underscore this potential problem by placing flags on catalogued completeness and uncertainty estimates of any candidate has no simulated sources lying beyond its location in parameter space.

To reiterate, we aim to recover mappings that enable us to go from measured candidate quantities to estimates of completeness and uncertainties. We do this by comparing the results obtained using the simulated sources to first define the measurement bias, which translates an observed quantity to an intrinsic one, and then evaluate the corresponding uncertainties and completeness given the pipeline results for model UDGs with those intrinsic properties. 
Readers not interested in further details of our completeness and error estimation can skip ahead to  \S\ref{sec:catalog}.

With the general approach defined, we focus on four  distinct aspects: the optimal window size, the number of bins, the parameter limits used for modeling, and the order of the polynomial model. These considerations and more detailed descriptions of our approaches for estimating parameter uncertainties and sample completeness are discussed separately below.

\begin{deluxetable}{lc}
\caption{Simulation Parameter Ranges for Uncertainties}
\tablewidth{\linewidth}
\label{tab:uncertain_params}
\tablehead{
\colhead{Parameter}&
\colhead{Distribution}\\
}
\startdata
$n$&[0.1,3.0)\\
$b/a$&[0.25,1.0)\\
$\theta$&[$-90^\circ,+90^\circ$)\\
$r_e$&[3.5\arcsec,20\arcsec)\\
$\mu_{0,g}$&[$22.5$ mag arcsec$^{-2}$, $27.5$ mag arcsec$^{-2}$)\\
$g-r$&[$0.0$  mag ,  $0.96$  mag]\\
$r-z$&[$-0.07$  mag,  $0.60$  mag] \\
\enddata
\tablecomments{The range given for $\mu_{0,g}$ is for the one of three processing runs with the largest range.  See text for details.}
\end{deluxetable}

\paragraph{Uncertainties}
Error estimates produced by GALFIT, derived from covariance matrices, are statistical in nature \citep{peng} and have been shown to significantly underestimate the true errors \citep{Haussler}.   We attempt to better define both any systematic error (parameter bias) and random uncertainties (the confidence limits) using our simulated sources.
The parameter error for a given simulated object is defined as the difference between the derived GALFIT value and the input value (GALFIT $-$ input). Because the errors are generally asymmetric, we define the bias as the median difference and the ``1$\sigma$" confidence limits as the 15.1 and 84.9 percentiles of the distribution for a set of simulated objects. 
We will present both the uncorrected measurements and our estimated parameter biases in the catalog. Users of the catalog are encouraged to apply the bias values (by subtracting the values presented in the catalog from the corresponding uncorrected measurement), but we present them separately for transparency and in the interest of reproducibility. 

For each of the ten parameters of interest
($\mu_{0,g}$, $\mu_{0,r}$, $\mu_{0,z}$, $m_g$, $m_r$, $m_z$
$r_e$, $b/a$, $n$, and $\theta$) we 
fit models to the uncertainties 
that are functions of four parameters. For the uncertainties in the structural components these include $\mu_{0,g}$, $r_e$, $n$, and $b/a$.  Uncertainties in the photometric components are modeled using $r_e$, $n$, and $b/a$ and the corresponding central surface brightness or magnitude of the band under evaluation.
We determine best fit models for the bias in each parameter and for the 15.1\% and 84.9\% confidence bounds for each parameter. 

The models are fit to the results obtained using a set of simulated sources. To define the sets, binning is required and we begin by exploring different values for the window size for that binning.  To simplify the problem, we  set the window size for each of the four parameters to be the same, yet undefined, fraction of each parameter's range. We then randomly place those windows to define a 4-D parameter bin. Note that we will define a large number of partially overlapping bins. We do this to mitigate sensitivity in the fitting to the exact placement of independent bins.
The parameter values defining each particular bin are the median values of those sources in the sample that lie within the bin. As such, fewer windows mean lower resolution and less coverage near the edge of the parameter ranges. Uncertainty estimates require a statistical sample to evaluate and, therefore, we require that windows contain at least 10 detections. This criterion leads to a further reduction in the range of parameter space that is well-modeled.  We evaluate window sizes that range from one-tenth to one quarter of the parameter ranges by laying down $10^6$ windows for each size and testing with an 7\textsuperscript{th} order polynomial model.  Two metrics are used when making our selections: the  Coefficients of Determination (R$^2$) of the models and the number of our 226 Stripe 82 UDGs whose GALFIT results fall out of the model's parameter space. These are both estimated as the averages of ten separate generations of the model.  

In our first attempts, we found that the lower confidence limits of the surface brightness were among the most difficult to model robustly.
Therefore, we pay special attention to the $\mu_{0,g}$ lower confidence limits to guide our selections. 
Based on the results of this exploration and considerations of fidelity toward the ends of the parameter range, we select a base window size of one-seventh the parameter range. 

Using this window size, we next explore the effects of the number of random bins on our metrics by testing data sets ranging from 100,000 to 3,000,000 samples. 
Although there is a slight improvement in R$^2$ above 2.5$\times 10^6$ points, we select this value because of the long computation times required to process significantly larger data sets when modeling high-order polynomials. The absence of significant differences between the R$^2$ metric obtained for the training and test sets indicates that overfitting is not a problem with data sets of this size. 
We find that parameter limits of $\mu_{0,g} \ge 23.3$ mag arcsec$^{-2}$, $r_e \ge 5.3^{\prime\prime}$, $b/a \ge 0.34$, and $n < 2.0$ provide the best compromise between maximizing R$^2$ and minimizing the number of UDGs falling outside of the model parameter space for our Stripe 82 UDGs.

We train separate models using all 2.5$\times10^6$ random bins for each of the 30 values to be estimated (lower confidence limits, median and upper confidence limits for each of 10 constrained parameters). The Coefficients of Determination, R$^2$, for the models are shown in Table \ref{tab:Rsq}. The vast majority are close to one suggesting that our selected models accurately predict errors and biases for those data sets. Except as noted below for $\theta$, the models provide excellent  approximations within their associated ranges.

With these models, we now calculate the expected systematic and random errors in each of the parameters anywhere within the parameter space. The biases, as well as the `1$\sigma$' confidence limits, are presented in the catalog. As shown in Table \ref{tab:Rsq}, the R$^2$ value for the median  of $\theta$, while clearly nonzero, is significantly smaller than those of the other models.  As expected, 
the error in $\theta$ 
as a function of all four parameters used for modeling ($\mu_{0,g}$, $r_e$, $b/a$, and $n$) is highly symmetric around zero, indicating a negligible bias in our $\theta$ determinations. To avoid adding noise, we set all $\theta$ biases to zero in the catalog.

\begin{deluxetable}{lccc}
\caption{Coefficients of Determination$^a$}
\label{tab:Rsq}
\tablehead{
\colhead{Parameter}&
\colhead{15.1\% C.I.}&
\colhead{Median}&
\colhead{84.9\% C.I.}
}
\startdata
$\mu_{0,g}$&0.806&0.965&0.964\\
$\mu_{0,r}$&0.791&0.965&0.964\\
$\mu_{0,z}$&0.832&0.974&0.974\\
$r_e$&0.988&0.975&0.918\\
$b/a$&0.955&0.794&0.929\\
$n$&0.988&0.973&0.850\\
$\theta$&0.960&0.406&0.958\\
$m_g$&0.911&0.980&0.973\\
$m_r$&0.908&0.981&0.972\\
$m_z$&0.951&0.986&0.981
\enddata
\tablenote{Each entry is the R$^2$ value obtained from the best fit model of that quantity. For example, the uppermost left entry represents the results for the model fit to recover the 15.1\% confidence interval for $\mu_{0,g}$.}
\end{deluxetable}

We conclude this section by commenting on subtleties that we have so far neglected. Our simulated sources are modeled as smooth UDGs, like those found in the Coma cluster \citep[e.g.,][]{vdk15a}.  Field UDGs often show much more structure \citep[e.g.,][]{rs, karunakaran} and the effect of this structure on parameter uncertainties is unclear. Modeling these effects is obviously much more complex because of the unconstrained nature of the range of possible structures. Moreover, our test samples for modeling are drawn from the same population as the training samples, and, therefore we cannot assume that the lack of overfitting also applies to real UDGs with significant morphological differences.
Despite these caveats, the recovery of simulated sources does provide baseline estimates of the uncertainties and completeness (see below for the latter). Ultimately, both the uncertainties and completeness will be validated through comparison with other surveys that have different selection, such as H{\small I} surveys, and structural analyses using other observations that are less sensitive to star formation, such as near-IR images. 

\paragraph{Completeness}
We define completeness as the probability that a UDG candidate with given structural and photometric parameters will be identified as such after passing through our entire pipeline, including the automated classification. We use four modeled parameters ($\mu_{0,g}$, $r_e$, $b/a$, and $n$) when assessing completeness and only include surviving simulations that meet the threshold limits defined by our model. Criteria for selecting these limits are discussed below. 

As we did for the uncertainty modeling, we explore different values for the window size and the degree of the polynomial model that we use in the fitting to define our final model.
To specify the model, we fit to completeness values defined for bins that span the parameter space. Within each bin, we calculate the probability of recovering a UDG candidate by evaluating the ratio of the number of simulated sources recovered by our pipeline processing to the number input.

We evaluate window sizes ranging from one-tenth to one quarter of the parameter ranges. We again lay down $10^6$ windows for each selected window size, but now require a minimum recovery fraction of 0.0004 to avoid spurious detections in unpopulated parts of parameter space and we test with a 7\textsuperscript{th} order polynomial model. At this point we set the model parameter limits to be $\mu_{0,g}$ $\ge$ 23 mag arcsec$^{-2}$, 4.0\arcsec $\le$ $r_e$ $<$ 20.0\arcsec, $b/a$ $\ge$ 0.30, and $n$ $<$ 2.4. 
$R^2$ values are near one for all window sizes. We select a window size of one-sixth of the parameter ranges to minimize the number of UDGs (3) falling out of the parameter space. Because the initial parameter limits used for evaluating window sizes produced excellent results, we see no need for further exploration of these limits and use them for further evaluations and modeling.

We now use the selected window size and parameter limits
to evaluate polynomial models of order ranging from 3 to 10.
We elect to use a 3\textsuperscript{rd} degree polynomial in order to minimize extrapolation errors.
With these choices, we have a model that provides us with estimates of the completeness at the specific location in parameter space for each of our candidates 
and we include these values in our catalog (\S \ref{sec:catalog}).

We show the sensitivity of the completeness fraction on candidate parameters in Figure \ref{fig:contours}. The completeness function is clearly complex and shows at least some sensitivity to each of the parameters we track.  The dominant factor appears to be central surface brightness, perhaps with S\'ersic index being the next most important, but the necessity of modeling it across at least this set of parameters is evident. 
Although \cite{vdb17} used a lower effective radius threshold, our completeness estimates show similar qualitative behavior to their estimates
in an effective radius vs. central surface brightness parameter space. 

Quantitative comparisons are difficult to make given the range of different choices and definitions in the two studies. For example, our maximum completeness
appears low, $< 0.5$, in Figure \ref{fig:contours} for two reasons. First, because a large fraction of our survey area is masked or rejected due to Galactic dust contamination, many  simulated sources land  in regions that are not analyzed and are therefore not recovered. Second, Figure \ref{fig:contours} shows the completeness marginalized over all of the unplotted parameters. As such, if our model suite includes many simulated sources of extremely faint surface brightness, then the completeness as presented in the Figure will appear to be very low. In practice, however, the completeness correction factor applied to any given  candidate is appropriate for its own multi-parameter characterization, and is therefore unaffected by the many much lower surface brightness objects we included in our modeling and failed to recover.

We now discuss two issues that potentially affect the interpretation of our completeness estimates. First, the survey has variable depth across its footprint because certain regions, such as Stripe 82, are imaged more frequently than others. The Legacy Surveys were constructed only to meet a minimum depth requirement, not to be homogeneous (although in practice most of the area is covered to roughly the same depth). As such, our sensitivity will vary across the sky. This phenomenon is captured by our random placement of simulations, assuming large-scale density fluctuations are negligible (e.g., assuming Virgo does not lie in an unusually deep region of the survey). However, it does lead to large scatter in completeness (a candidate with $\mu_{0,g} = 27$  mag arcsec$^{-2}$ may be detected in one region of the sky, but not in another), which means that our completeness estimates are representative for the whole survey and should not be applied to small, limited regions. Second, as mentioned above, our simulations are modeled as simple S\'ersic profiles while the structure of real UDGs can be more complex and, therefore, have different completeness distributions.  Ongoing work (Karunakaran et al., in prep) involving a comparison of H{\small I}-selected and optically-selected UDGs will, in part, address questions regarding our completeness for such sources.

\begin{figure*}[ht]
\includegraphics[width=1.0\textwidth]{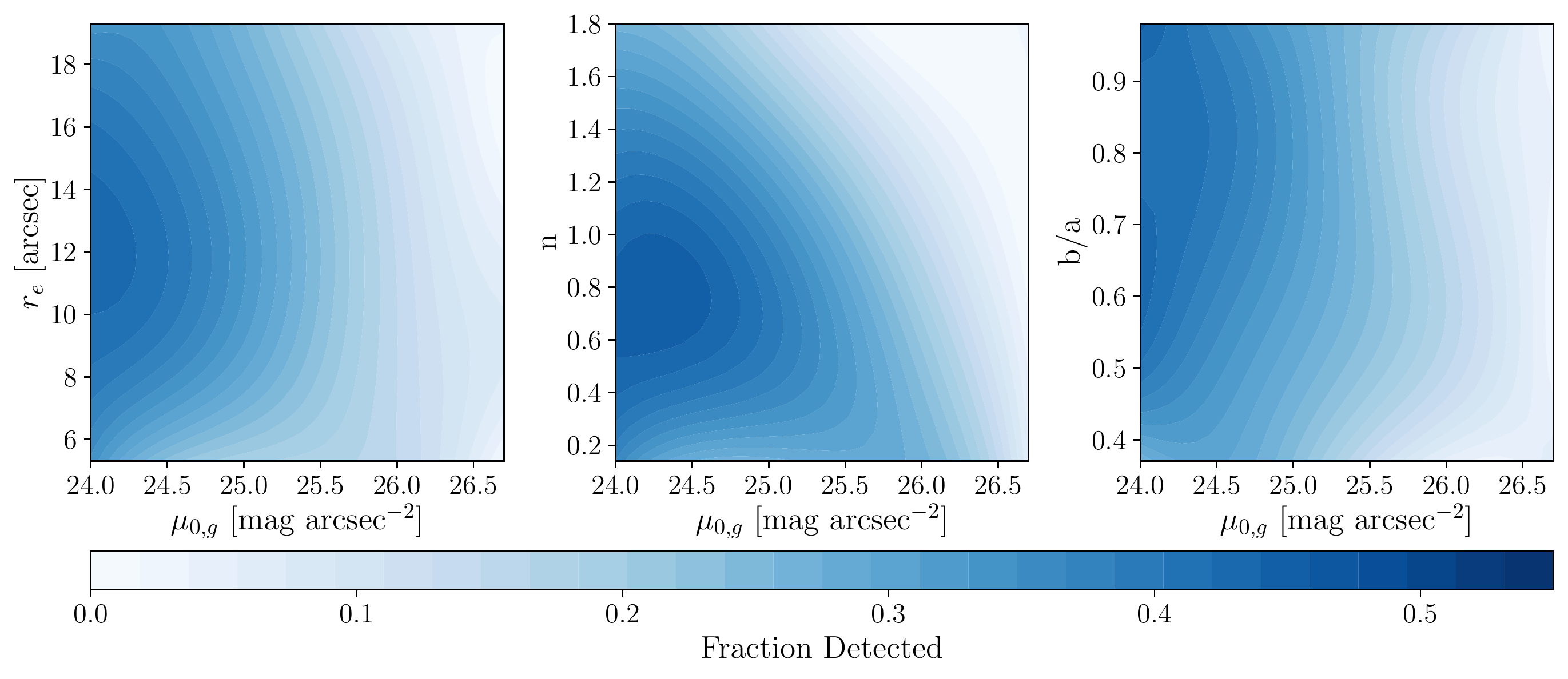}
\caption{The fractional completeness as a function of candidate UDG parameters.  The structural parameters ($r_e$, n, and b/a) are shown on the vertical axes to demonstrate their effects when coupled with central surface brightness, $\mu_{0,g}$. We have similar models for the $r$ and $z$ band data. As plotted, the maximum completeness appears to be $<$ 0.5 but for certain types of candidates the completeness is much larger (see text for discussion).}
\label{fig:contours}
\end{figure*}

\section{The Catalog}
\label{sec:catalog}

We present our catalog of 226 UDG candidates within the Stripe 82 region, with a description of the catalog entries presented in Table \ref{tab:catalog} (and the full catalog in the electronic version of the Table). Each parameter entry includes its GALFIT estimate as well as the bias and confidence limits produced by our models.  We additionally flag any entry where we had to extrapolate the fitted model beyond the range of the constraints.  Uncertainties and completeness estimates for flagged entries are suspect and should be used with caution.  Photometric parameters are not corrected for extinction, but extinction values are included in the table for those who wish to use them. Our extinction estimates (A$_g$, A$_r$, A$_z$) are calculated using the SDSS $g$, $r$, and $z$ coefficients in Table 6 of \cite{Schlafly} for an R$_V$ = A$_V$/E(B-V)$_{SFD}$ value of 3.1. E(B-V)$_{SFD}$ is estimated using the  dustmaps.py \citep{green} SFD dust map based on \cite{SFD}. Alternatively, users may wish to apply different extinction determinations. Parameters are corrected for bias before their completeness values are estimated.  Those with bias corrections considered unreliable (completeness flag $\ne 0$) have flag $= 2$, to distinguish them from objects with flag $= 1$ where the completeness correction was derived beyond the model parameter range. Our approach inherently prevents us from including values approaching the $b/a$ physical limit of 1.0. Therefore, when estimating completeness for our catalog UDGs we set any $b/a$ value $>$ 0.97 to 0.97 which occupies a region relatively insensitive to this parameter (Figure \ref{fig:contours}). 

We show the distribution in the sky of the 226 candidates relative to known ``normal" galaxies that satisfy $1000 < cz/{\rm km\ sec}^{-1} < 15000$ and $12 < m < 18 $ in Figure \ref{fig:distribution}. There are two aspects of note. First, the distribution of candidate UDGs tends to follow the higher density regions among the normal galaxies. This association was also seen in Paper I and qualitatively confirms that UDGs are correlated with normal galaxies. Second, the projected density of candidate UDGs falls steeply for $RA < -15^\circ$ and $RA > 40^\circ$. These declines mirror the increase in Galactic dust, and hence excluded regions, evident in Figure \ref{fig:DustFP}. 

\begin{deluxetable*}{lrr}
\caption{The Catalog$^a$}
\label{tab:catalog}
\tablehead{
\colhead{Column Name}&
\colhead{Description}&
\colhead{Format}\\
}
\startdata
SMDG&Object Name&SMDG designator plus coordinates\\
RA&Right Ascension (J2000.0)&decimal degrees\\
Dec&Declination (J2000.0)&decimal degrees\\
r\_e&effective radius&angular (arcsec)\\
r\_e\_upper\_uncertainty&effective radius 1$\sigma$ upper uncertainty&angular (arcsec)\\
r\_e\_bias&effective radius measurement bias&angular (arcsec)\\
r\_e\_lower\_uncertainty&effective radius  1$\sigma$ lower uncertainty&angular (arcsec)\\
r\_e\_flag&effective radius uncertainty model flag&0 = good, 1 = extrapolated\\
AR&axis ratio ($b/a$)&unitless\\
AR\_upper\_uncertainty&axis ratio 1$\sigma$ upper uncertainty&unitless\\
AR\_bias&axis ratio measurement bias&unitless\\
AR\_lower\_uncertainty&axis ratio 1$\sigma$ lower uncertainty&unitless\\
AR\_flag&axis ratio uncertainty model flag&0 = good, 1 = extrapolated\\
n&S\'ersic index&unitless\\
n\_upper\_uncertainty&S\'ersic index 1$\sigma$ upper uncertainty&unitless\\
n\_bias&S\'ersic index measurement bias&unitless\\
n\_lower\_uncertainty&S\'ersic index 1$\sigma$ lower uncertainty&unitless\\
n\_flag&S\'ersic index uncertainty model flag&0 = good, 1 = extrapolated\\
PA&major axis position angle&defined to be [$-$90,90) measured \\
&&N to E, in degrees\\
PA\_upper\_uncertainty&major axis position angle 1$\sigma$ upper uncertainty&degrees\\
PA\_bias&major axis position angle measurement bias&degrees\\
PA\_lower\_uncertainty&major axis position angle 1$\sigma$ lower uncertainty&degrees\\
PA\_flag&major axis position angle uncertainty model flag&0 = good, 1 = extrapolated\\
mu0\_$X$&central surface brightness in band $X$ ($X \equiv$ g,r,z)&AB mag arcsec$^2$\\
mu0\_$X$\_upper\_uncertainty&central surface brightness 1$\sigma$ upper uncertainty in band $X$&AB mag arcsec$^2$\\
mu0\_$X$\_bias& central surface brightness measurement bias in band $X$&AB mag arcsec$^2$\\
mu0\_$X$\_lower\_uncertainty&central surface brightness 1$\sigma$ lower uncertainty in band $X$&AB mag arcsec$^2$\\
mu0\_$X$\_flag&central surface brightness uncertainty model flag in band $X$&0 = good, 1 = extrapolated\\
mag\_$X$&total apparent magnitude in band $X$&AB mag\\
mag\_$X$\_upper\_uncertainty&total apparent magnitude 1$\sigma$ upper uncertainty in band $X$&AB mag\\
mag\_$X$\_bias&total apparent magnitude measurement bias in band $X$&AB mag\\
mag\_$X$\_lower\_uncertainty&total apparent magnitude 1$\sigma$ lower uncertainty in band $X$&AB mag\\
mag\_$X$\_flag&total apparent magnitude uncertainty model flag in band $X$&0 = good, 1 = extrapolated\\
SFD&Optical depth at SMDG location from \cite{schlegel}&unitless\\
A\_$X$&Corresponding extinction at SMDG location in band $X$&AB mag\\
Comp&Fractional completeness for similar UDGs&unitless\\
Comp\_flag&Completeness model flag&0 = good, 1=extrapolated, \\
&&2=biases extrapolated\\
\enddata
\tablenote{The catalog is available as the electronic version of this Table.}
\end{deluxetable*}

\begin{figure*}
\includegraphics[width=1.0\textwidth]{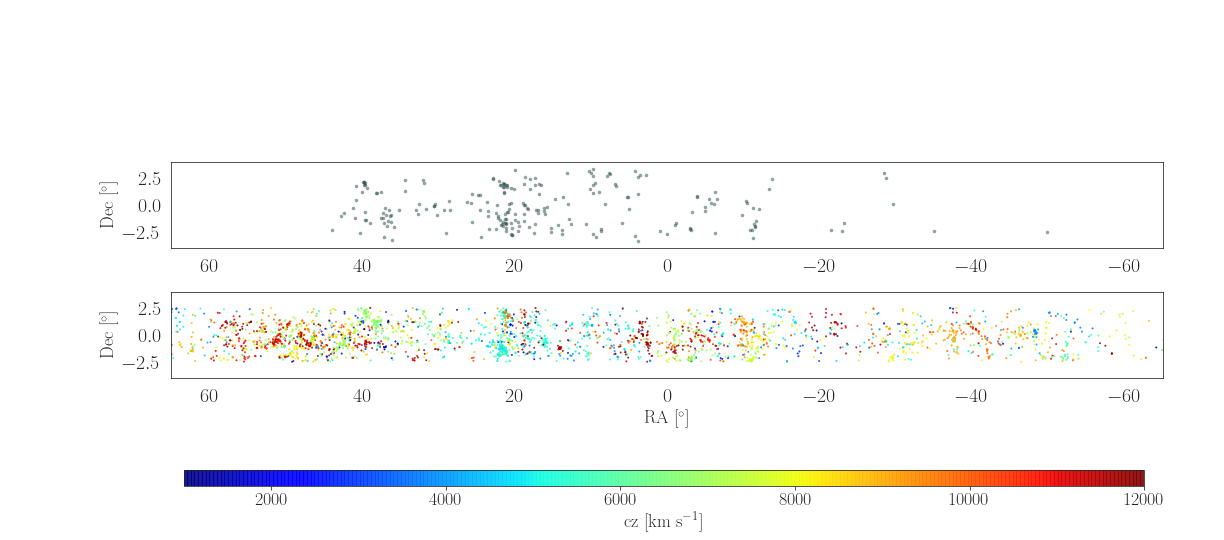}
\caption{Comparison of the distribution on the sky of UDG candidates in the Stripe 82 region (upper panel) and a set of normal galaxies ($1000 < cz < 15000\,\mathrm{km\,s^{-1}}$ and $12 < m < 18$) from NED that consists mostly of SDSS galaxies (lower panel). The vertical and horizontal scales are different, leading to an artificial stretching of coherent physical structures in the vertical direction. For the normal galaxies, which have available redshifts, we color code by $cz$ as illustrated by the color bar. The most identifiable structures in the UDG candidate distribution correspond to normal galaxy concentrations at $cz \sim 5500$ km sec$^{-1}$.}
\label{fig:distribution}
\end{figure*}

\subsection{Comparison to Previous UDG Searches}

The overlapping regions between the area we have surveyed and previous surveys provide both a check of our results but also clear examples of how the various searches differ.

\subsubsection{\cite{roman17a, roman17b}}

\cite{roman17a} compiled a sample of 80 UDG candidates in an area corresponding to 8$\times 8$ Mpc around Abell 168 (z = 0.045) using the Stripe 82 SDSS data to explore the role of environment on UDGs. This cluster is about twice as far as the Coma cluster (z = 0.023), meaning that our angular criterion used to select galaxies with $r_e > 2.5$ kpc at Coma will select galaxies with $r_e \gtrsim 5$ kpc at the distance of Abell 168. Such galaxies are rare, so we do not expect a large sample. Among the 80 candidates \cite{roman17a} identified and measured are two with $r_e > 5^{\prime\prime}$ (IAC37 and IAC63). Our catalog contains both (SMDG0115352+001434 and SMDG0113177-001417). The remainder of their sample lies below our angular size cut, demonstrating how different UDG studies can probe distinct populations by virtue of their selection criteria.

\cite{roman17b} present data for 11 UDGs in Hickson Compact Groups within Stripe 82. Of those 11, 5 have $r_e < 5$\arcsec\ and 3 others have $\mu_{0,g} < 24$ mag arcsec$^{-2}$. Given these properties, only 3 could potentially be included in our catalog. Among those 3, two have $\mu_{0,g} = 24$ mag arcsec$^{-2}$ in the \citet{roman17b} catalog, just at the bright end of our central surface brightness boundary. We do not have matches to either of these (UDG-B1 and UDG-B2) and, indeed, 
in our estimation these two galaxies (SMDG0320211-011014 and
SMDG0038239+010621) have $\mu_{0,g} = 23.37$ and 23.48 mag arcsec$^{-2}$, respectively, and so fail to satisfy our criterion. Our catalog does include the remaining one (UDG-R4). Again, this comparison is a cautionary tale about the nature of different UDG samples. We recover the galaxies that are well within our selection criteria, but uncertainties can move objects in and out of samples when their parameters are near the selection boundaries. Simulations to understand the uncertainties and related completeness levels are critical.

\subsubsection{\cite{leisman}}
\label{sec:leisman}

\cite{leisman} identified a set of large, low surface brightness galaxies from H{\small I} detections. Although their sources are spread over a much larger area of sky than what we currently present, there are five galaxies in their catalog that both overlap our survey area and satisfy our criteria for $\mu_{0,g}$ and $r_e$ (AGC numbers 102375, 103435, 100288, 322019, and 334349). The first  three are in our catalog, while the last two are not. The last two, however, lie at RA $< -10$, where Galactic dust is significantly affecting our completeness, and would not affect the H{\small I} detection, and furthermore AGC~334349 has a measured surface brightness, 24.1 mag arcsec$^{-2}$ in the \cite{leisman} compilation, near our selection limit. Going back through our results, we find that both galaxies are in regions of high optical depth ($\tau \sim 0.1$), and hence in excluded regions, and also that we measure $\mu_{0,g} = 23.95$ mag arcsec$^{-2}$ for ACG~334349, which would make it slightly brighter than our cutoff. Therefore, there are no unexplained differences in the catalogs. This partly addresses a concern that our procedure might be biased against star-forming, clumpy, systems. 

\subsubsection{\cite{greco}}

\cite{greco} present a large set of low surface brightness galaxies identified in the Wide Layer of the Hyper Suprime-Cam Subaru Strategic Program \citep{aihara}. That survey has fields distributed around the sky, but one field happens to have partial overlap with our Stripe 82 survey region. A comparison to that catalog offers an opportunity to explore how different selection criteria applied to higher resolution images obtained with a larger telescope result in a different catalog. Unfortunately, of the 781 objects in their catalog, only 139 also satisfy our central surface and angular size criterion, and of those only 8 overlap with our catalog. Of those 8, we recover two (objects 191 and 201 in their catalog). We do not recover their objects 197, 203, 210, 216, 230, and 250. Object 210, with $r_e = 5.13^{\prime\prime}$, does not match our criteria. Objects 197, 216, 230, and 250 did not pass our $r_e$ or $\mu_{0,g}$ criteria in early stages of processing. The last three are close to our selection boundaries in the parametric values provided by \cite{greco}, suggesting random scatter could easily move them outside the boundaries. Finally, object 203 was detected in many individual images, but the centroid disagreed sufficiently among the various images that we did not reach a critical number of coincident detections. Statistically, we estimate that this level of misalignment can be expected in 1 out 500 of our detected sources. We will attempt to mitigate this problem in future applications of the pipeline, but do not address it here because, statistically, we do not expect any more such issues for the Stripe 82 sample.
Our completeness calculation captures these various issues because the simulated sources are processed with the same pipeline. We conclude, given that we do detect the objects and understand why they are not included in our catalog, that we are sensitive to galaxies such as those presented by \cite{greco}.

\subsubsection{\cite{tanoglidis}}
\label{sec:tanoglidis}

\cite{tanoglidis} present what is probably the closest comparable effort to that described here. They analyze data from the Dark Energy Survey \citep{des}, which are included in the Legacy Surveys that we analyze, and also employ machine learning tools to classify objects. We find that 112 out of the 226 candidates we present were also cataloged by \cite{tanoglidis}. The nature of the 114 candidates that are not matched to objects in their catalog is clarified in Figure \ref{fig:tanoglidis}. The matching is mostly complete for $\mu_{0,g} < 25$ mag arcsec$^{-2}$, about 50\% for 25 mag arcsec$^{-2} < \mu_{0,g} < 25.5$ mag arcsec$^{-2}$, and almost entire incomplete for $\mu_{0,g} > 25.5$ mag arcsec$^{-2}$. \cite{tanoglidis} claim to be $> 50\%$ complete for $\bar{\mu}_{e,r} = 26.0$ mag arcsec$^{-2}$, which given the differences between $\mu_0$ and $\bar{\mu}_e$, a typical $g-r$ color, and our own incompleteness, is an estimate that appears to be in line with the results in Figure \ref{fig:tanoglidis}.

\begin{figure}[ht]
\centering{
\includegraphics[width=0.45\textwidth]{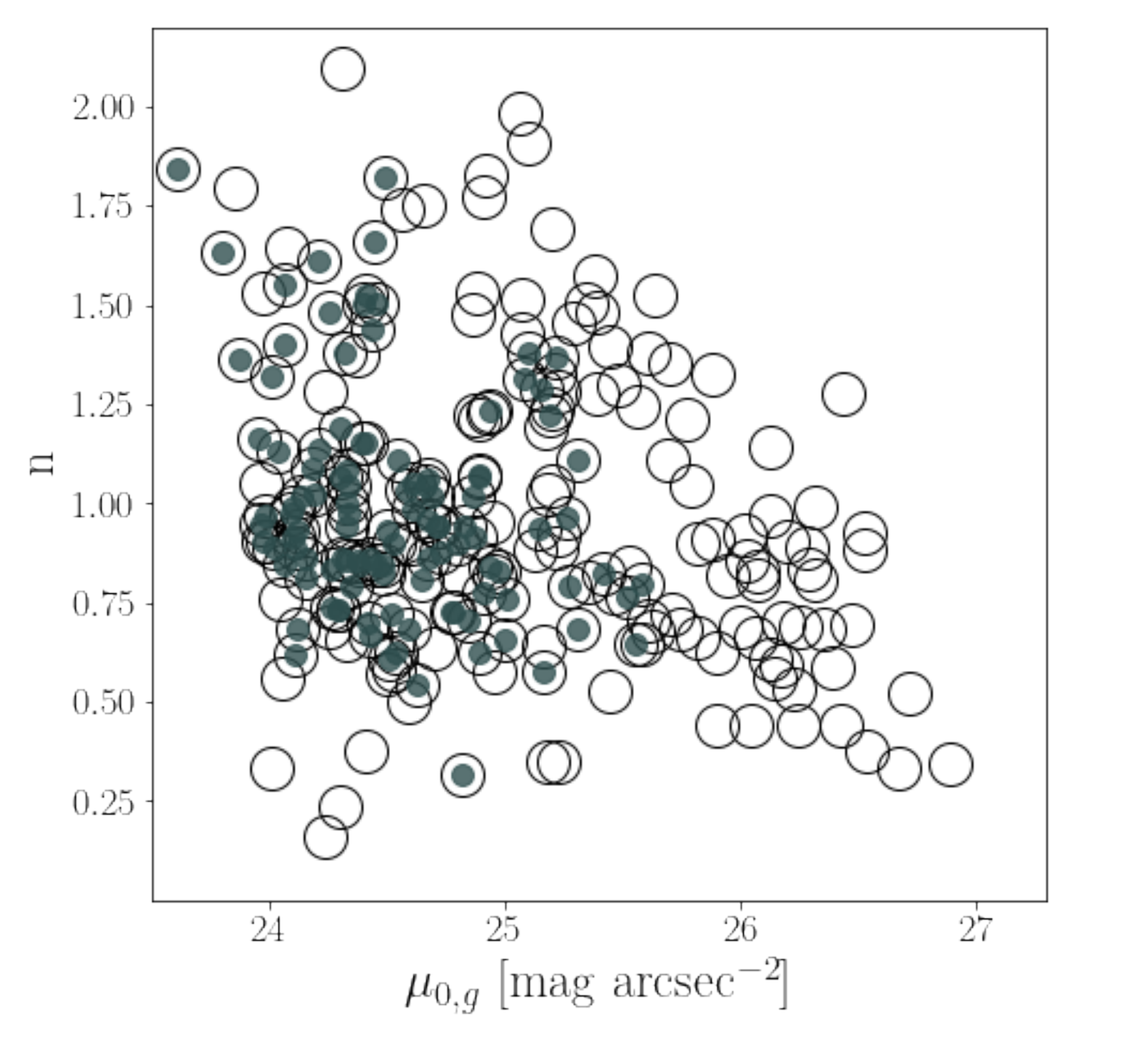}
\caption{Distribution of UDG candidates in the $\mu_{0,g}-n$ space. The open circles represent the 226 objects in our catalog and the filled sources represent the objects in our catalog for which we found a match within 5\arcsec\ in the \cite{tanoglidis} compilation. }
\label{fig:tanoglidis}}
\end{figure}

It is important to note that different surveys have different strengths and do not necessarily supersede one another. For example, although SMUDGes is about one  magnitude deeper in central surface brightness sensitivity, the \cite{tanoglidis} catalog includes objects of smaller angular extent than SMUDGes and therefore has a much larger projected surface density of objects. Among the reasons one might prefer to use the SMUDGes catalog, aside from the fainter surface brightness limit, are the detailed completeness calculations, the quantitative treatment of Cirrus contamination, which becomes more important as one progresses to fainter surface brightness, and (eventually) the dual hemisphere coverage and larger survey footprint.

\section{Properties of UDG Candidates}
\label{sec:discussion}

We now present some preliminary discussion of the properties of our candidates. Again, one must bear in mind that the fitted models are idealized and may lead to systematic errors in the derived properties for candidates that deviate significantly from the assumed model (e.g., those candidates that have very asymmetric surface brightness distributions). The questions raised here will benefit greatly from the full SMUDGes survey, which we estimate will contain roughly 30$\times$ as many candidates. As such, while we present and discuss some results, we limit the discussion and interpretation. Here we are examining only the Stripe 82 candidates. The Coma catalog from Paper I was compiled in a sufficiently different manner, without completeness and full uncertainty estimates, that we choose not to include it here. The Coma area will be reprocessed, along with the rest of the Legacy Surveys, using the procedure described here and presented elsewhere. We apply the estimated bias to all of the measured parameters, and consider only those systems that have an unflagged completeness correction estimate. 

\begin{figure}[ht]
\centering{
\includegraphics[width=0.45\textwidth]{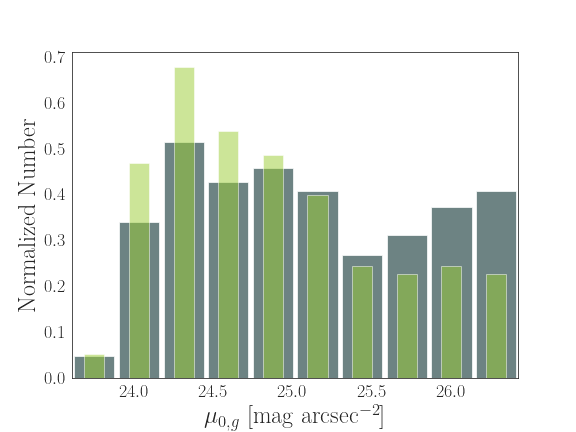}
\caption{The distribution of central surface brightness. The relative distributions (normalized histograms) for the cataloged objects (narrow bars, light green) is to be compared to the completeness-corrected distribution (wide bars, dark grey). What was a originally a steeply declining distribution with increasing (fainter) central surface brightness, dropping by $\sim$ a factor of two across the range, becomes a flatter distribution, where the decline is at most 20\%. There are a small number of objects at $\mu_{0,g}<24.0$ because the bias correction made them brighter than our selection criterion.}
\label{fig:surface_brightness}}
\end{figure}

\subsection{Distances and Physical Sizes}

Measuring distances, with which one can then assess the physical characteristics of the candidates, is the most challenging aspect of compiling a UDG catalog. Indeed, most UDGs catalogs, like this one, are in reality UDG-candidate catalogs. Although efforts to build up samples of spectroscopically confirmed UDGs are ongoing \citep{vdk15b,kadowaki17,alabi,chilingarian,karunakaran,kadowaki21}, the largest samples of UDGs with physical characteristics come from ``distance-by-association". In such an approach one associates UDGs in redshift with a large overdensity of galaxies, the Coma cluster being the most striking example \citep{vdk15a}. This interpretation, for Coma in particular, has been confirmed with a number of spectroscopic redshifts \citep{vdk15b, kadowaki17, alabi}. The approach is used in larger surveys as well \citep[e.g.,][]{tanoglidis}.

Six of our candidates have available spectroscopic redshift measurements in the literature. To obtain these we searched the NASA Extragalactic Database (NED) and SIMBAD for projected matches within 5\arcsec. The obtained   recessional velocities are presented in Table \ref{tab:specz}. Physical effective radii are calculated assuming a smooth Hubble flow and our measurements of $r_e$.  There are a few aspects of note. First, the first system in the Table (SMDG0014598+023448) appears to have an unrealistically large value for $r_e$. This could indicate that the associated redshift is for a second galaxy seen in projection. Upon inspection, there is no such system within 5\arcsec\ of the source. The large apparent size could also indicate a tidal nature for the source. Tidal features are often identified as low surface brightness galaxy candidates and can have large extents \citep{kadowaki21}. This does not appear to be the case here either. We revisited the GALFIT modeling and that appears to be fine as well. It is possible that the redshift is incorrect. The H{\small I} detection from which the redshift is determined is described as weak by the authors of that study \citep{impey96}. Second, the last object in the Table (SMDG2334535+011226) has a small $r_e$ and does not meet the standard UDG size criterion of $r_e > 1.5$ kpc. This object illustrates how the selection on angular size alone can lead to a sample contaminated by non-UDG, low surface brightness galaxies. The remainder of the objects satisfy the UDG size criterion, and two of the objects (SMDG0032359+000652 and SMDG0058555+003627) are among the high end of the UDG size distribution. These two objects demonstrate that physically large UDGs are not exclusively found in extremely dense environments such as the Coma cluster.

This small sampling is consistent with, but not compelling evidence for, a potential claim that the majority of our candidates are, in fact, UDGs. To consider the issue further, note that our angular effective radius lower limit of 5.3$^{\prime\prime}$ corresponds to 1.5 kpc, the community standard UDG minimum size, for $cz \approx 4000$ km s$^{-1}$. As such, we could have a large population of nearby, physically small low surface brightness galaxies contaminating our sample. This ambiguity cannot be resolved for any specific candidate without spectroscopy, but we can investigate bulk sample properties statistically using the strength of the angular two-point cross correlation function between UDG candidates and normal galaxies for which we have redshifts, as done in Paper I. We divide the sample of SDSS normal galaxies out to $cz = 15,000$ km s$^{-1}$ into four equal redshift interval samples. The first bin represents the survey volume in which our candidates might not satisfy the UDG size criterion. The two-point cross correlation function is the weakest for this innermost bin and stronger, by over a factor of two, for the next bin in distance. The amplitudes decrease thereafter. Therefore, pending direct distance estimates, we conclude that although the sample is certainly not clean of contaminating non-UDGs, the majority of the sample consists of galaxies that satisfy the UDG size criterion.
This result is in part due to the fact that Stripe 82 mostly avoids dominant nearby structure. As such, our conclusion is not transferable to any arbitrary region within the full SMUDGes survey.

\begin{deluxetable}{lrr}
\caption{UDG Candidates in Stripe 82 with Spectroscopic Redshifts}
\label{tab:specz}
\tablehead{
\colhead{Name}&
\colhead{$r_e$}&
\colhead{$cz$}\\
&\colhead{[kpc]} &
\colhead{[km sec$^{-1}$]}\\
}
\startdata
SMDG0014598+023448&10.33$^{+0.31}_{-0.47}$&17941\\
SMDG0031454+024256&2.41$^{+0.31}_{-0.89}$&2382\\
SMDG0032359+000652&5.27$^{+0.07}_{-0.03}$&13165\\
SMDG0058555+003627&4.45$^{+0.29}_{-0.25}$&5437\\
SMDG0107238+010008&3.04$^{+0.04}_{-0.02}$&4190\\
SMDG2334535+011226&1.15$^{+0.37}_{-0.28}$&1306\\
\enddata
\end{deluxetable}

\subsection{Surface Brightness Distribution}

The surface brightness distribution of these galaxies is intimately tied to the identification of these objects. Intuitively, we expect the completeness to decline as one considers galaxies with lower central surface brightnesses and this intuition is confirmed by our modeling (Figure \ref{fig:contours}). The effect of this trend can be seen in the comparison of the measured $\mu_{0,g}$ distribution and that corrected by our completeness estimates (Figure \ref{fig:surface_brightness}). While the raw distribution shows a significant decline (factor of $\sim$ 2) from the bright to the faint end of the survey range, the completeness-corrected version shows a much milder decline of perhaps no more than $\sim$ 20\%. Given the 2.5 mag range of our survey, even a flat distribution in the numbers of systems suggests a factor of $\sim$ 10 less stellar mass density contained in galaxies at the lower vs. the upper end of the range. The properties of the dark matter halos that these systems occupy is of evident interest, but becomes more and more difficult to ascertain as the surface brightness drops below even the limits of this survey. 

We have not reached a surface brightness for which there is a marked decline in the numbers of halos containing identifiable stellar populations. This finding is not entirely surprising given that in the Local Group, where one can reach far lower effective surface brightness by counting individual stars, there are known objects reaching a surface brightness of nearly 30 mag arcsec$^{-2}$ \citep{drlica-wagner}, 
although those surface brightnesses refer to the mean surface brightness rather than a central one and those systems are physically smaller galaxies than those that are the focus of SMUDGes. For various reasons, including more significant contamination from Galactic cirrus, it will become increasingly difficult to push to fainter central surface brightnesses, but upcoming surveys, such as the LSST with the Rubin Observatory, should be able to probe this regime.

\subsection{Joint Parameter Distributions}

Joint parameter distributions can help us understand the underlying behavior that defines individual parameter distributions. 

\subsubsection{Color vs. S\'ersic index}

Consider the joint distribution of $g-r$ color and the S\'ersic index, $n$, shown in Figure \ref{fig:joint_color_n}. In either the raw version or the completeness-corrected version, we find that the tail of objects with $n > 1$ are red ($g-r \gtrsim 0.45)$. The bluer objects are confined to the smallest values of $n$, even to somewhat lower $n$ values than the bulk of the red objects. The completeness-corrected version of the distribution begins to suggest that the $n>1$ red population may be a distinct population rather than simply a high-$n$ tail. The statistics are such that this feature may also simply be a random fluctuation in the joint parameter distribution. A larger sample is needed to establish this hypothesis, and that will be available from the complete SMUDGes survey. Finally, we caution that the interpretation of this result is complicated by the possibilities of a correlation between $n$ and $r_e$, differences in the redshift and size distributions as a function of color, and systematics in fitting a smooth model to the clumpier blue galaxies that are unaccounted for in our uncertainty estimates.

\begin{figure}
\begin{center}
\includegraphics[angle=-90,width=0.45\textwidth]{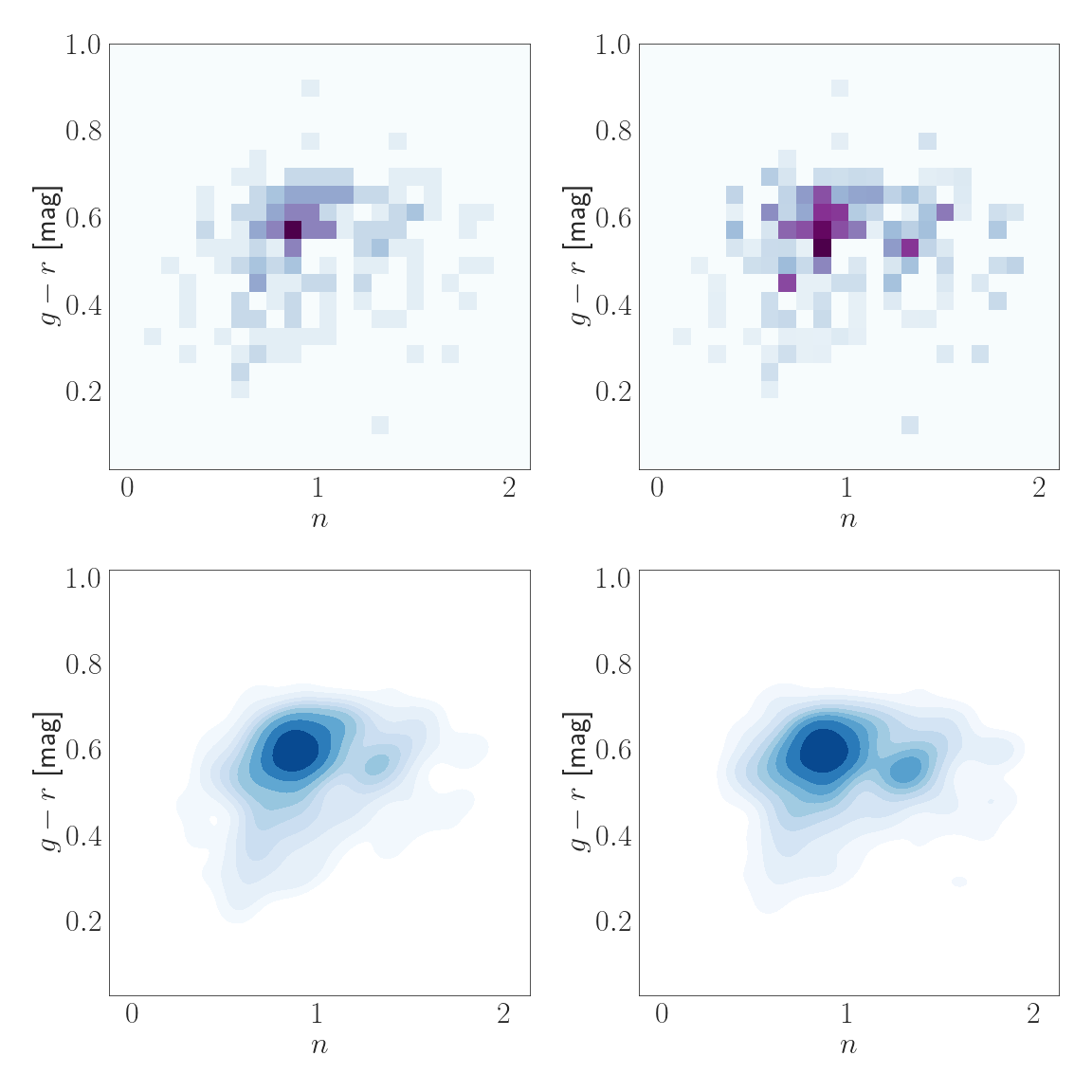}
\caption{The joint distribution of color, $g-r$, and S\'ersic index, $n$. The left set of panels show the distribution for the set of objects in the catalog, while the right set shows the distribution once we apply the 
completeness corrections. The lower panels show the same distributions as the corresponding upper panels in terms of contours.
}
\label{fig:joint_color_n}
\end{center}
\end{figure}

\subsubsection{S\'ersic Index vs. central surface brightness}

There is previously published evidence for a relation between $n$ and $\mu_0$ \citep{vdk15a,mancera,kadowaki21}. \cite{kadowaki21} find a highly statistically-significant correlation, such that $n$ decreases as the central surface brightness gets fainter, among spectroscopically-confirmed UDGs. We see a corresponding trend in Figure \ref{fig:joint_n_mu} if we ignore the population of  candidates at $n\sim 1.3$. Interestingly, this $n\sim 1.3$ population is not present among the spectroscopically-confimed UDG sample \citep{kadowaki21}, perhaps indicating that these are not UDGs, or that they are UDGs with a real structural difference (such as a nuclear star cluster) or that there is asymmetric uncertainty in the recovered parameter distribution. Visual examination of these high $n$ sources suggests a variety of explanations for many of them, including the presence of nuclear star clusters, superimposed background sources, and a suggestion for a higher concentration second component, like a bulge, in some systems. Discriminating among these possibilities is beyond the scope of this paper, but will be treated in subsequent work. We close here by noting that if lower surface brightness UDGs do indeed also have lower $n$, this makes them doubly difficult to detect. 

\begin{figure}[ht]
\centering{
\includegraphics[angle=-90,width=0.45\textwidth]{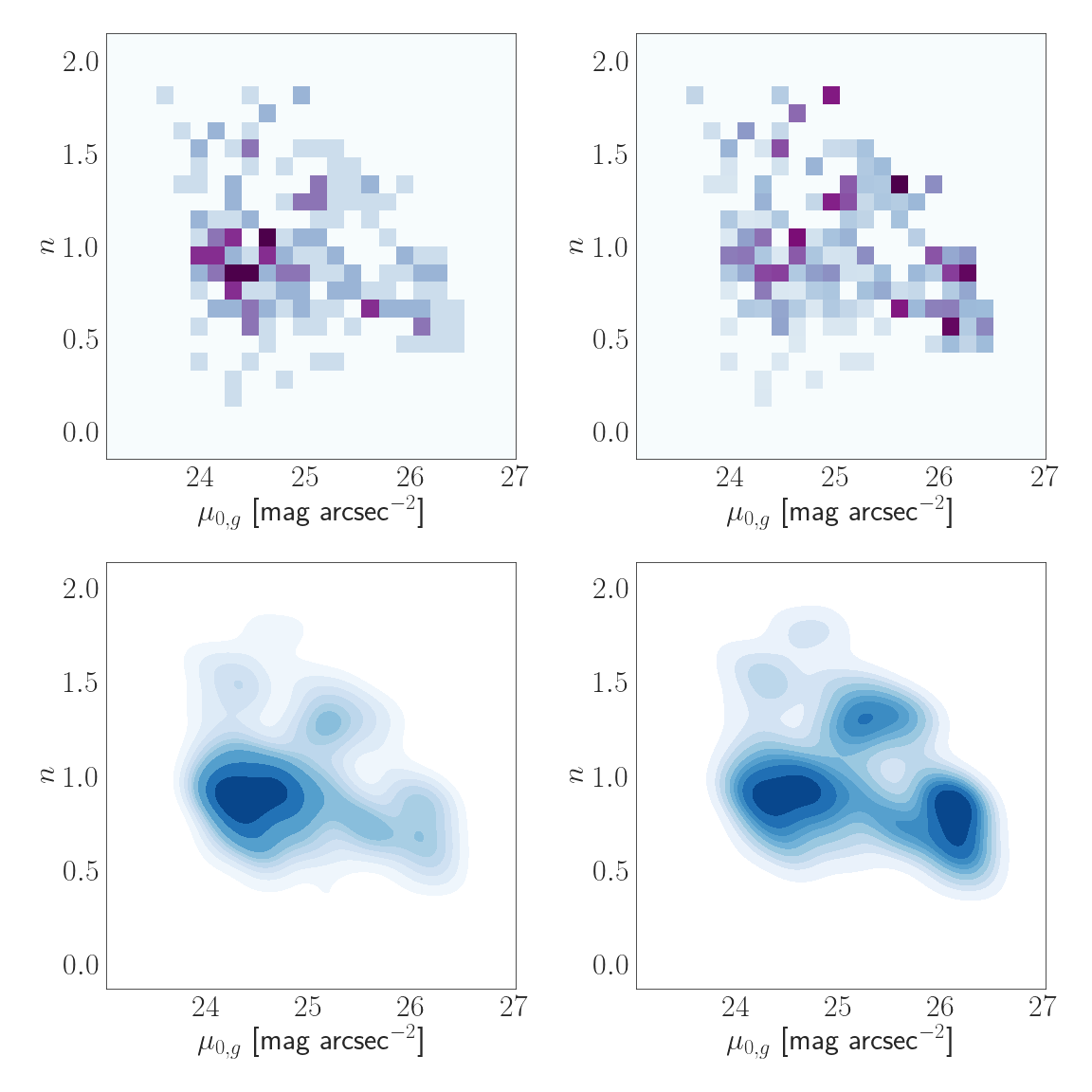}
\caption{The joint distribution of S\'ersic index, $n$, and central surface brightness, $\mu_{0,g}$. The left set of panels show the distribution for the set of objects in the catalog, while the right set shows the distribution once we apply the completeness corrections. The lower panels show the same distributions as the corresponding upper panels in terms of contours.
}
\label{fig:joint_n_mu}}
\end{figure}

\subsubsection{Color vs. central surface brightness}

A third joint parameter distribution that provides some interesting insights is that of color and central surface brightness (Figure \ref{fig:joint_color_mu}). There are three aspects that we find noteworthy. First, the completeness correction clearly emphasizes the lowest surface brightness systems, which as we described before, leads to a much less dramatic decline in the number of candidates with declining surface brightness. Second, there is a suggestion in the data that among the red population the color becomes bluer as $\mu_{0,g}$ increases. For otherwise similar systems, an increase in $\mu_{0,g}$ will correspond to an increase in $m_g$ and ultimately to $M_g$ and $M_*$. Therefore, the trend in color may reflect the slope of the stellar mass-metallicity relation in this population. Confirmation of this interpretation awaits distance estimates for a significant fraction of these candidates but would be in line with what has been found otherwise for UDGs \citep{barbosa}. Finally, and perhaps most intriguingly, is the absence of low surface brightness blue candidates. The blue candidates are limited to $\mu_{0,g} < 25$ mag arcsec$^{-2}$. In other words, had UDGs been defined to have $\mu_{0,g}>25$ mag arcsec$^{-2}$ they would have been a nearly entirely red population. This is not a result of incompleteness because the completeness corrections make no qualitative difference to the distribution of blue galaxies in this Figure. Furthermore, we have shown that we 
did not miss any H{\small I}-detected sources \citep{leisman} in the Stripe 82 region (\S\ref{sec:leisman}) and the result is qualitatively similar to that found by \cite{greco}.

While the blue UDGs are expected to fade as their star formation stops and they age, thereby populating the red part of these diagrams, the absence of fainter blue systems suggests that what appears to be a continuing red sequence may end not far below our current surface brightness limit. 
Taking the 100 SMUDGes sources that also have S-PLUS data for which the spectral energy distributions were analyzed \citep{barbosa}, we can now ask how the blue galaxies among that set would evolve if star formation stopped. In Figure \ref{fig:ageing} we compare the current $g-$band central surface brightness with what we calculate it would be in 5 Gyr. To calculate the fading, we adopt the stellar population results from \cite{barbosa} and allow each galaxy to age passively for an additional 5 Gyr. On average, $\mu_{0,g}$ increases by about 0.9 mag, but the full distribution of sources matches well the range of $\mu_{0,g}$ that we find for our red galaxies. We conclude that fading of the blue candidates we find can explain the fainter red candidates that we find, and that the absence of fainter blue candidates, which we were capable of finding, suggests that the red candidate population is not expected to extend significantly fainter than $\mu_{0,g} = 27$ mag arcsec$^{-2}$, unless those form via a separate formation channel.

\begin{figure}[ht]
\centering{
\includegraphics[angle=-90,width=0.45\textwidth]{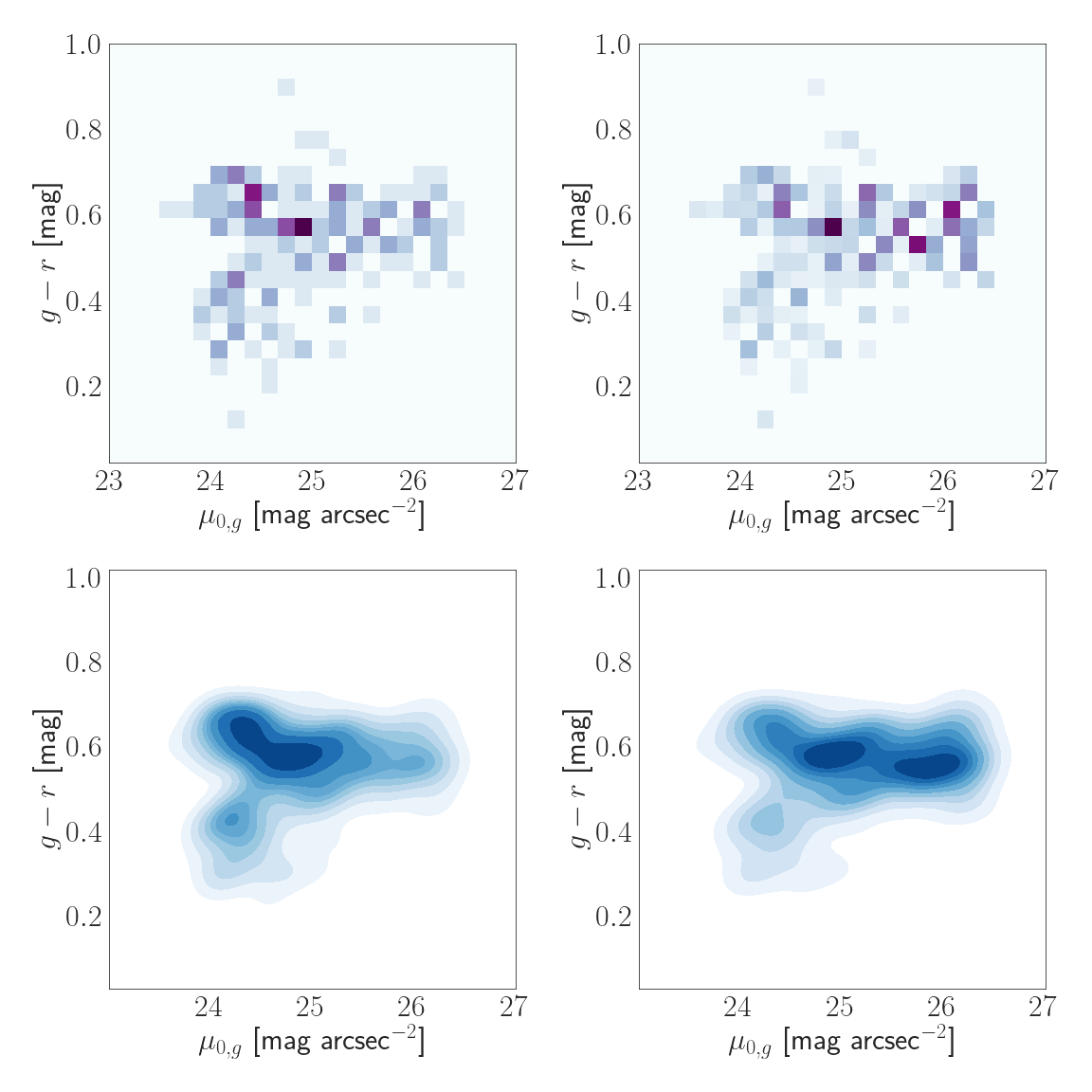}
\caption{The joint distribution of color, $g-r$, and central surface brightness, $\mu_{0,g}$. The left set of panels show the distribution for the set of objects in the catalog, while the right set shows the distribution once we apply the completeness corrections. The lower panels show the same distributions as the corresponding upper panels in terms of contours.
}
\label{fig:joint_color_mu}}
\end{figure}

\begin{figure}[ht]
\centering{
\includegraphics[width=0.45\textwidth]{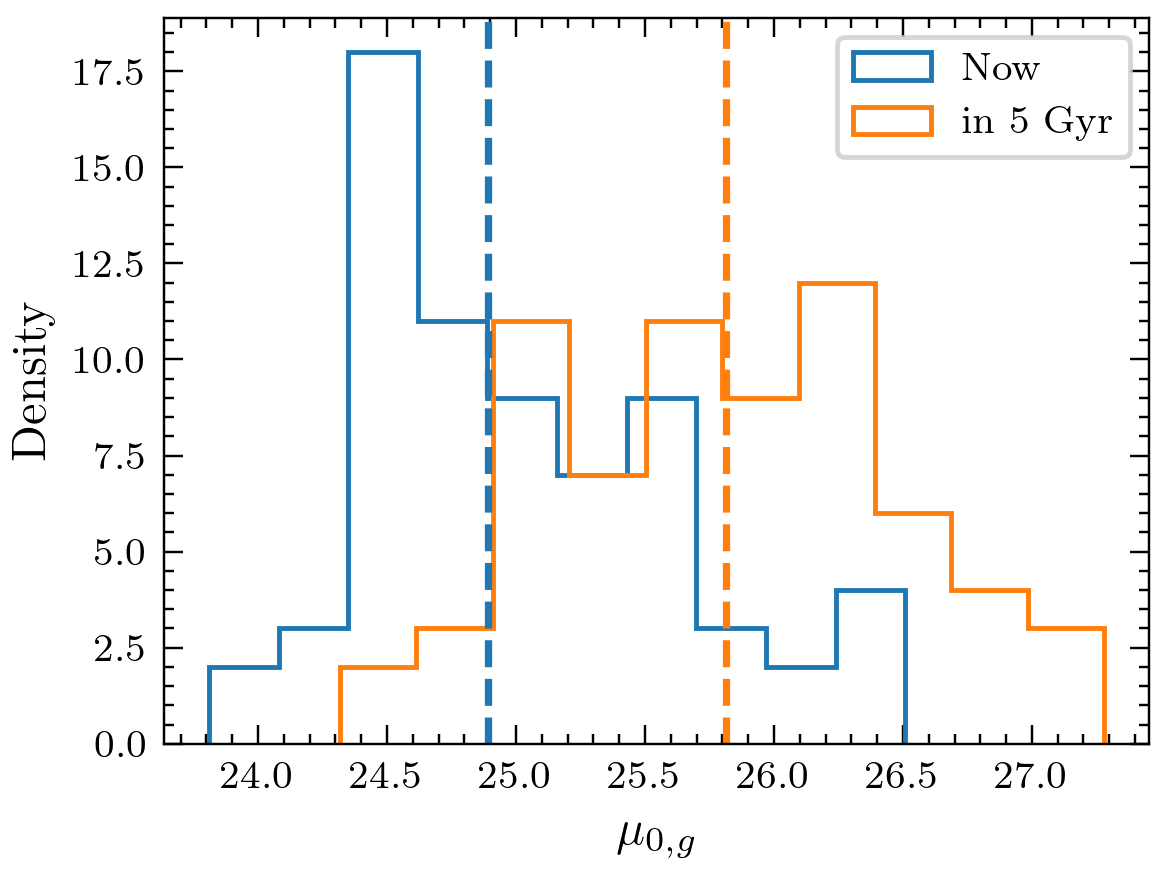}
\caption{The $g-$band central surface brightness distribution of blue UDG candidates now and in 5 Gyr, assuming no star formation from the current time. On average, $\mu_{0,g}$ increases by 0.9 mag arcsec$^{-2}$ and the distribution matches well the range we measure for the red UDG candidate population.
}
\label{fig:ageing}}
\end{figure}

\subsection{Axis Ratios vs. Color}

We compare the axis ratio distribution of red and blue candidates, with the color boundary drawn at $g-r = 0.45$ mag, in Figure \ref{fig:axis_ratios}. The distribution of the red candidates is peaked at $b/a \sim 0.75$ while that of the blue candidates is  a bit more 'flat-topped', with relative deficit to the red distribution at the largest values of $b/a$. This qualitative result remains whether or not we apply the completeness-correction. A two-sided Kolmogorov-Smirnov test of the completeness-corrected distributions indicates that the two distributions are different with 93\% confidence.
Pending higher statistical significance results, understanding how contamination may vary with color, and an exploration of internal dynamical heating in UDGs,
the differences in the distributions may preclude the simplest models in which all of the red  candidates are simply faded, blue candidates. Further exploration of this issue might provide a strong constraint on formation models. 
A similar, but more thorough argument, for dwarf galaxies overall, is made on a range of morphological grounds by \cite{carlsten}.

\begin{figure}[ht]
\centering{
\includegraphics[width=0.45\textwidth]{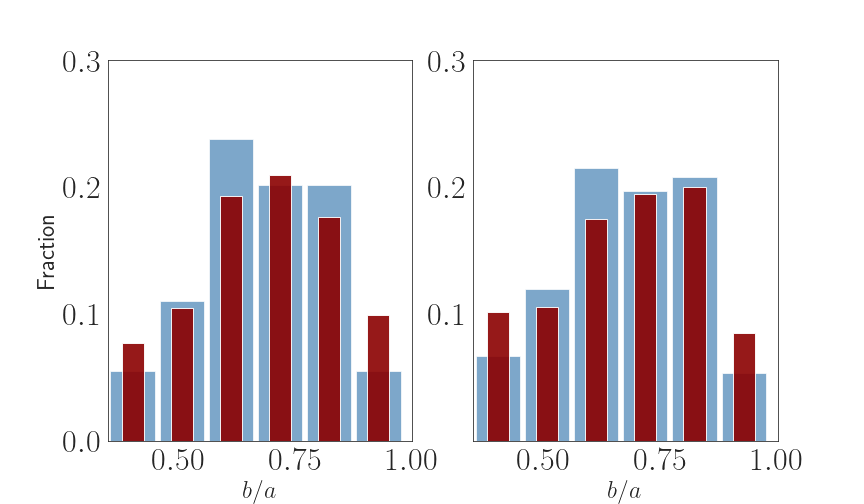}
\caption{Distribution of axis ratios for blue (wider, blue bars) and red (narrower, red bars) UDG candidates. Left panel shows the raw distribution and the right panel the completeness-corrected distribution. The distributions have been normalized to enable more direct comparison. The blue sample contains 47 candidates and the red sample contains 158.}
\label{fig:axis_ratios}}
\end{figure}

\section{Summary}

We present 226 ultra-diffuse galaxy candidates ($r_e > 5.3$\arcsec, $\mu_{0,g} > 24$ mag arcsec$^{-2}$) in Stripe 82 that are recovered using our improved procedure developed in anticipation of processing the entire Legacy Surveys footprint. We have implemented a variety of improvements that make both material improvements in the output catalog and strengthen both the robustness and efficiency of the algorithms. We expect the description presented here of our procedure, in combination with that presented in \cite{smudges}, will suffice for the full catalogs to be presented in subsequent releases. 

The advancements relative to Paper I described here include:

\smallskip
\noindent
1) fitting that includes the effects of the point-spread function with a floating S\'ersic index structural parameter ($n$), which leads to a wide range of best-fitting $n$, but a clear preference for $n < 1$, and allows us to explore possible relationships between $n$ and other parameters; 

\smallskip
\noindent
2) an expanded wavelet filtering criteria that enables us to simultaneously search for candidates of larger angular extent;

\smallskip
\noindent
3) careful consideration of Galactic dust and the false detections it can give rise to, leading to a selection criterion based on comparison of {\sl Planck}$-$ and {\sl WISE}$-$based estimates of the dust column density;

\smallskip
\noindent
4) parametric uncertainty estimates and completeness estimates that are based on full pipeline processing of millions of simulated sources that span the full parameter range; and

\smallskip
\noindent
5) further refinements and training of our automated candidate classification. 

We compare our catalog to previous catalogs \citep{roman17a,roman17b,leisman,tanoglidis}, and find that we either recover the published sources in the overlapping parameter ranges or understand the cause for the discrepancy. We have a sensitivity $\sim$1.5 mag fainter in central surface brightness than the largest previous catalog of this region \citep{tanoglidis}. Our recovery of the \cite{leisman} sources is also important because those were selected as H{\small I} sources, indicating that we are sensitive to the H{\small I}-rich UDG population as well.

We present a set of findings based on this preliminary sample that do not depend on the (unknown) distance to the candidate:

\smallskip
\noindent
1) after correcting for incompleteness, there is no significant decline in the number of UDG candidates as a function of $g-$band central surface brightness down to the limit of our survey ($\sim$ 26.5 mag arcsec$^{-2}$); 

\smallskip
\noindent
2) S\'ersic structural parameter $n$ and $g-r$ color correlate such that bluer galaxies have smaller measured $n$;

\smallskip
\noindent
3) lower surface brightness UDG candidates tend to have smaller $n$, which confirms previous results by \cite{vdk15a,mancera,kadowaki21};

\smallskip
\noindent
4) the bulk of blue ($g-r < 0.45$ mag) UDG candidates have central surface brightness $\mu_{0,g} \lesssim 25$ mag arcsec$^{-2}$ and can fade to match the UDG red sequence we observe down to $\mu_{0,g} \sim 26.5$ mag arcsec$^{-2}$, but will not fade sufficiently to match any UDGs that may lie well below our surface brightness sensitivity; and 

\smallskip
\noindent
5) the differing axis ratio distributions for the red and blue UDG candidates suggests that, despite the population modeling described in (4), the red UDG candidates are not just simply faded blue candidates, but instead also require internally- or externally-driven morphological transformation. 

We look forward to significant upcoming improvements. First, the processing of the entire Legacy Surveys data will yield a final SMUDGes sample that contains $\sim$ 30$\times$ as many candidates. Second, the key to making progress in this field is having the distance estimates necessary to determining physical quantities. Spectroscopic redshifts are coming from a variety of efforts \citep{kadowaki21,karunakaran}, and given the large footprint and size of SMUDGes there are many opportunities for distance-by-association, a technique that connects UDGs with local overdensities of known galaxies \citep{vdk15a,tanoglidis}. Our initial efforts, which will be described elsewhere, suggest that we may be able to provide distance-by-association distance estimates for as much as 20\% of the SMUDGes catalog. Lastly, deeper ground and space based surveys will both provide improved measurements of these UDG candidates and uncover lower surface brightness ones.

\begin{acknowledgments}

DZ, RD, JK, and HZ acknowledge financial support from NSF AST-1713841 and AST-2006785. AD's research is supported by NOIRLab. KS acknowledges funding from the Natural Sciences and Engineering Research Council of Canada (NSERC). An allocation of computer time from the UA Research Computing High Performance Computing (HPC) at the University of Arizona and the prompt assistance of the associated computer support group is gratefully acknowledged. We thank the anonymous referee for comments that improved the clarity of the presentation.
 
This research has made use of the NASA/IPAC Extragalactic Database (NED), which is operated by the Jet Propulsion Laboratory, California Institute of Technology, under contract with NASA. 

This research depends directly on images from the Dark Energy Camera Legacy Survey (DECaLS; Proposal ID 2014B-0404; PIs: David Schlegel and Arjun Dey). Full acknowledgment at \url{https://www.legacysurvey.org/acknowledgment/}.

\end{acknowledgments}

\facilities{Blanco}

\software{
Astropy              \citep{astropy1, astropy2},
astroquery           \citep{astroquery},
dustmaps             \cite{green},
GALFIT               \citep{peng},
keras                \citep{keras},
lmfit                \citep{newville},
Matplotlib           \citep{matplotlib},
NumPy                \citep{numpy},
pandas               \citep{pandas},
sep                  \citep{sep},
Scikit-learn         \citep{sklearn},
SciPy                \citep{scipy1, scipy2},
Source Extractor     \citep{bertin},
SWarp                \citep{Swarp}}

\bibliography{references.bib}{}

\begin{thebibliography}{}
\expandafter\ifx\csname natexlab\endcsname\relax\def\natexlab#1{#1}\fi
\providecommand{\url}[1]{\href{#1}{#1}}
\providecommand{\dodoi}[1]{doi:~\href{http://doi.org/#1}{\nolinkurl{#1}}}
\providecommand{\doeprint}[1]{\href{http://ascl.net/#1}{\nolinkurl{http://ascl.net/#1}}}
\providecommand{\doarXiv}[1]{\href{https://arxiv.org/abs/#1}{\nolinkurl{https://arxiv.org/abs/#1}}}

\bibitem[{{Abbott} {et~al.}(2018){Abbott}, {Abdalla}, {Allam}, {Amara},
  {Annis}, {Asorey}, {Avila}, {Ballester}, {Banerji}, {Barkhouse}, {Baruah},
  {Baumer}, {Bechtol}, {Becker}, {Benoit-L{\'e}vy}, {Bernstein}, {Bertin},
  {Blazek}, {Bocquet}, {Brooks}, {Brout}, {Buckley-Geer}, {Burke}, {Busti},
  {Campisano}, {Cardiel-Sas}, {Carnero Rosell}, {Carrasco Kind}, {Carretero},
  {Castander}, {Cawthon}, {Chang}, {Chen}, {Conselice}, {Costa}, {Crocce},
  {Cunha}, {D'Andrea}, {da Costa}, {Das}, {Daues}, {Davis}, {Davis}, {De
  Vicente}, {DePoy}, {DeRose}, {Desai}, {Diehl}, {Dietrich}, {Dodelson},
  {Doel}, {Drlica-Wagner}, {Eifler}, {Elliott}, {Evrard}, {Farahi}, {Fausti
  Neto}, {Fernandez}, {Finley}, {Flaugher}, {Foley}, {Fosalba}, {Friedel},
  {Frieman}, {Garc{\'\i}a-Bellido}, {Gaztanaga}, {Gerdes}, {Giannantonio},
  {Gill}, {Glazebrook}, {Goldstein}, {Gower}, {Gruen}, {Gruendl}, {Gschwend},
  {Gupta}, {Gutierrez}, {Hamilton}, {Hartley}, {Hinton}, {Hislop}, {Hollowood},
  {Honscheid}, {Hoyle}, {Huterer}, {Jain}, {James}, {Jeltema}, {Johnson},
  {Johnson}, {Kacprzak}, {Kent}, {Khullar}, {Klein}, {Kovacs}, {Koziol},
  {Krause}, {Kremin}, {Kron}, {Kuehn}, {Kuhlmann}, {Kuropatkin}, {Lahav},
  {Lasker}, {Li}, {Li}, {Liddle}, {Lima}, {Lin}, {L{\'o}pez-Reyes}, {MacCrann},
  {Maia}, {Maloney}, {Manera}, {March}, {Marriner}, {Marshall}, {Martini},
  {McClintock}, {McKay}, {McMahon}, {Melchior}, {Menanteau}, {Miller},
  {Miquel}, {Mohr}, {Morganson}, {Mould}, {Neilsen}, {Nichol}, {Nogueira},
  {Nord}, {Nugent}, {Nunes}, {Ogand o}, {Old}, {Pace}, {Palmese},
  {Paz-Chinch{\'o}n}, {Peiris}, {Percival}, {Petravick}, {Plazas}, {Poh},
  {Pond}, {Porredon}, {Pujol}, {Refregier}, {Reil}, {Ricker}, {Rollins},
  {Romer}, {Roodman}, {Rooney}, {Ross}, {Rykoff}, {Sako}, {Sanchez}, {Sanchez},
  {Santiago}, {Saro}, {Scarpine}, {Scolnic}, {Serrano}, {Sevilla-Noarbe},
  {Sheldon}, {Shipp}, {Silveira}, {Smith}, {Smith}, {Smith}, {Soares-Santos},
  {Sobreira}, {Song}, {Stebbins}, {Suchyta}, {Sullivan}, {Swanson}, {Tarle},
  {Thaler}, {Thomas}, {Thomas}, {Troxel}, {Tucker}, {Vikram}, {Vivas},
  {Walker}, {Wechsler}, {Weller}, {Wester}, {Wolf}, {Wu}, {Yanny}, {Zenteno},
  {Zhang}, {Zuntz}, {DES Collaboration}, {Juneau}, {Fitzpatrick}, {Nikutta},
  {Nidever}, {Olsen}, {Scott}, \& {NOAO Data Lab}}]{des}
{Abbott}, T.~M.~C., {Abdalla}, F.~B., {Allam}, S., {et~al.} 2018, \apjs, 239,
  18, \dodoi{10.3847/1538-4365/aae9f0}

\bibitem[{{Aihara} {et~al.}(2018){Aihara}, {Arimoto}, {Armstrong}, {Arnouts},
  {Bahcall}, {Bickerton}, {Bosch}, {Bundy}, {Capak}, {Chan}, {Chiba}, {Coupon},
  {Egami}, {Enoki}, {Finet}, {Fujimori}, {Fujimoto}, {Furusawa}, {Furusawa},
  {Goto}, {Goulding}, {Greco}, {Greene}, {Gunn}, {Hamana}, {Harikane},
  {Hashimoto}, {Hattori}, {Hayashi}, {Hayashi}, {He{\l}miniak}, {Higuchi},
  {Hikage}, {Ho}, {Hsieh}, {Huang}, {Huang}, {Ikeda}, {Imanishi}, {Inoue},
  {Iwasawa}, {Iwata}, {Jaelani}, {Jian}, {Kamata}, {Karoji}, {Kashikawa},
  {Katayama}, {Kawanomoto}, {Kayo}, {Koda}, {Koike}, {Kojima}, {Komiyama},
  {Konno}, {Koshida}, {Koyama}, {Kusakabe}, {Leauthaud}, {Lee}, {Lin}, {Lin},
  {Lupton}, {Mandelbaum}, {Matsuoka}, {Medezinski}, {Mineo}, {Miyama},
  {Miyatake}, {Miyazaki}, {Momose}, {More}, {More}, {Moritani}, {Moriya},
  {Morokuma}, {Mukae}, {Murata}, {Murayama}, {Nagao}, {Nakata}, {Niida},
  {Niikura}, {Nishizawa}, {Obuchi}, {Oguri}, {Oishi}, {Okabe}, {Okamoto},
  {Okura}, {Ono}, {Onodera}, {Onoue}, {Osato}, {Ouchi}, {Price}, {Pyo}, {Sako},
  {Sawicki}, {Shibuya}, {Shimasaku}, {Shimono}, {Shirasaki}, {Silverman},
  {Simet}, {Speagle}, {Spergel}, {Strauss}, {Sugahara}, {Sugiyama}, {Suto},
  {Suyu}, {Suzuki}, {Tait}, {Takada}, {Takata}, {Tamura}, {Tanaka}, {Tanaka},
  {Tanaka}, {Tanaka}, {Terai}, {Terashima}, {Toba}, {Tominaga}, {Toshikawa},
  {Turner}, {Uchida}, {Uchiyama}, {Umetsu}, {Uraguchi}, {Urata}, {Usuda},
  {Utsumi}, {Wang}, {Wang}, {Wong}, {Yabe}, {Yamada}, {Yamanoi}, {Yasuda},
  {Yeh}, {Yonehara}, \& {Yuma}}]{aihara}
{Aihara}, H., {Arimoto}, N., {Armstrong}, R., {et~al.} 2018, \pasj, 70, S4,
  \dodoi{10.1093/pasj/psx066}

\bibitem[{{Alabi} {et~al.}(2018){Alabi}, {Ferr{\'e}-Mateu}, {Romanowsky},
  {Brodie}, {Forbes}, {Wasserman}, {Bellstedt}, {Mart{\'\i}n-Navarro},
  {Pandya}, {Stone}, \& {Okabe}}]{alabi}
{Alabi}, A., {Ferr{\'e}-Mateu}, A., {Romanowsky}, A.~J., {et~al.} 2018, \mnras,
  479, 3308, \dodoi{10.1093/mnras/sty1616}

\bibitem[{{Amorisco} \& {Loeb}(2016)}]{amorisco16}
{Amorisco}, N.~C., \& {Loeb}, A. 2016, \mnras, 459, L51,
  \dodoi{10.1093/mnrasl/slw055}

\bibitem[{{Annis} {et~al.}(2014){Annis}, {Soares-Santos}, {Strauss}, {Becker},
  {Dodelson}, {Fan}, {Gunn}, {Hao}, {Ivezi{\'c}}, {Jester}, {Jiang},
  {Johnston}, {Kubo}, {Lampeitl}, {Lin}, {Lupton}, {Miknaitis}, {Seo}, {Simet},
  \& {Yanny}}]{annis}
{Annis}, J., {Soares-Santos}, M., {Strauss}, M.~A., {et~al.} 2014, The
  Astrophysical Journal, 794, 120, \dodoi{10.1088/0004-637X/794/2/120}

\bibitem[{{Astropy Collaboration} {et~al.}(2013){Astropy Collaboration},
  {Robitaille}, {Tollerud}, {Greenfield}, {Droettboom}, {Bray}, {Aldcroft},
  {Davis}, {Ginsburg}, {Price-Whelan}, {Kerzendorf}, {Conley}, {Crighton},
  {Barbary}, {Muna}, {Ferguson}, {Grollier}, {Parikh}, {Nair}, {Unther},
  {Deil}, {Woillez}, {Conseil}, {Kramer}, {Turner}, {Singer}, {Fox}, {Weaver},
  {Zabalza}, {Edwards}, {Azalee Bostroem}, {Burke}, {Casey}, {Crawford},
  {Dencheva}, {Ely}, {Jenness}, {Labrie}, {Lim}, {Pierfederici}, {Pontzen},
  {Ptak}, {Refsdal}, {Servillat}, \& {Streicher}}]{astropy1}
{Astropy Collaboration}, {Robitaille}, T.~P., {Tollerud}, E.~J., {et~al.} 2013,
  \aap, 558, A33, \dodoi{10.1051/0004-6361/201322068}

\bibitem[{{Astropy Collaboration} {et~al.}(2018){Astropy Collaboration},
  {Price-Whelan}, {Sip{\H{o}}cz}, {G{\"u}nther}, {Lim}, {Crawford}, {Conseil},
  {Shupe}, {Craig}, {Dencheva}, {Ginsburg}, {VanderPlas}, {Bradley},
  {P{\'e}rez-Su{\'a}rez}, {de Val-Borro}, {Aldcroft}, {Cruz}, {Robitaille},
  {Tollerud}, {Ardelean}, {Babej}, {Bach}, {Bachetti}, {Bakanov}, {Bamford},
  {Barentsen}, {Barmby}, {Baumbach}, {Berry}, {Biscani}, {Boquien}, {Bostroem},
  {Bouma}, {Brammer}, {Bray}, {Breytenbach}, {Buddelmeijer}, {Burke},
  {Calderone}, {Cano Rodr{\'\i}guez}, {Cara}, {Cardoso}, {Cheedella}, {Copin},
  {Corrales}, {Crichton}, {D'Avella}, {Deil}, {Depagne}, {Dietrich}, {Donath},
  {Droettboom}, {Earl}, {Erben}, {Fabbro}, {Ferreira}, {Finethy}, {Fox},
  {Garrison}, {Gibbons}, {Goldstein}, {Gommers}, {Greco}, {Greenfield},
  {Groener}, {Grollier}, {Hagen}, {Hirst}, {Homeier}, {Horton}, {Hosseinzadeh},
  {Hu}, {Hunkeler}, {Ivezi{\'c}}, {Jain}, {Jenness}, {Kanarek}, {Kendrew},
  {Kern}, {Kerzendorf}, {Khvalko}, {King}, {Kirkby}, {Kulkarni}, {Kumar},
  {Lee}, {Lenz}, {Littlefair}, {Ma}, {Macleod}, {Mastropietro}, {McCully},
  {Montagnac}, {Morris}, {Mueller}, {Mumford}, {Muna}, {Murphy}, {Nelson},
  {Nguyen}, {Ninan}, {N{\"o}the}, {Ogaz}, {Oh}, {Parejko}, {Parley}, {Pascual},
  {Patil}, {Patil}, {Plunkett}, {Prochaska}, {Rastogi}, {Reddy Janga},
  {Sabater}, {Sakurikar}, {Seifert}, {Sherbert}, {Sherwood-Taylor}, {Shih},
  {Sick}, {Silbiger}, {Singanamalla}, {Singer}, {Sladen}, {Sooley},
  {Sornarajah}, {Streicher}, {Teuben}, {Thomas}, {Tremblay}, {Turner},
  {Terr{\'o}n}, {van Kerkwijk}, {de la Vega}, {Watkins}, {Weaver}, {Whitmore},
  {Woillez}, {Zabalza}, \& {Astropy Contributors}}]{astropy2}
{Astropy Collaboration}, {Price-Whelan}, A.~M., {Sip{\H{o}}cz}, B.~M., {et~al.}
  2018, \aj, 156, 123, \dodoi{10.3847/1538-3881/aabc4f}

\bibitem[{Barbary(2016)}]{sep}
Barbary, K. 2016, {SEP: Source Extractor as a library},
  \dodoi{10.21105/joss.00058}

\bibitem[{{Barbosa} {et~al.}(2020){Barbosa}, {Zaritsky}, {Donnerstein},
  {Zhang}, {Dey}, {Mendes de Oliveira}, {Sampedro}, {Molino}, {Costa-Duarte},
  {Coelho}, {Cortesi}, {Herpich}, {Hernandez-Jimenez}, {Santos-Silva},
  {Pereira}, {Werle}, {Overzier}, {Cid Fernandes}, {Smith Castelli}, {Ribeiro},
  {Schoenell}, \& {Kanaan}}]{barbosa}
{Barbosa}, C.~E., {Zaritsky}, D., {Donnerstein}, R., {et~al.} 2020, \apjs, 247,
  46, \dodoi{10.3847/1538-4365/ab7660}

\bibitem[{{Bennet} {et~al.}(2017){Bennet}, {Sand}, {Crnojevi{\'c}}, {Spekkens},
  {Zaritsky}, \& {Karunakaran}}]{bennet}
{Bennet}, P., {Sand}, D.~J., {Crnojevi{\'c}}, D., {et~al.} 2017, \apj, 850,
  109, \dodoi{10.3847/1538-4357/aa9180}

\bibitem[{{Bertin} \& {Arnouts}(1996)}]{bertin}
{Bertin}, E., \& {Arnouts}, S. 1996, \aaps, 117, 393,
  \dodoi{10.1051/aas:1996164}

\bibitem[{{Bertin} {et~al.}(2002){Bertin}, {Mellier}, {Radovich}, {Missonnier},
  {Didelon}, \& {Morin}}]{Swarp}
{Bertin}, E., {Mellier}, Y., {Radovich}, M., {et~al.} 2002, Astronomical
  Society of the Pacific Conference Series, Vol. 281, {The TERAPIX Pipeline},
  ed. D.~A. {Bohlender}, D.~{Durand}, \& T.~H. {Handley}, 228

\bibitem[{{Carleton} {et~al.}(2019){Carleton}, {Errani}, {Cooper},
  {Kaplinghat}, {Pe{\~n}arrubia}, \& {Guo}}]{carleton19}
{Carleton}, T., {Errani}, R., {Cooper}, M., {et~al.} 2019, \mnras, 485, 382,
  \dodoi{10.1093/mnras/stz383}

\bibitem[{{Carlsten} {et~al.}(2021){Carlsten}, {Greene}, {Greco}, {Beaton}, \&
  {Kado-Fong}}]{carlsten}
{Carlsten}, S.~G., {Greene}, J.~E., {Greco}, J.~P., {Beaton}, R.~L., \&
  {Kado-Fong}, E. 2021, arXiv e-prints, arXiv:2105.03435.
\newblock \doarXiv{2105.03435}

\bibitem[{{Chilingarian} {et~al.}(2019){Chilingarian}, {Afanasiev}, {Grishin},
  {Fabricant}, \& {Moran}}]{chilingarian}
{Chilingarian}, I.~V., {Afanasiev}, A.~V., {Grishin}, K.~A., {Fabricant}, D.,
  \& {Moran}, S. 2019, arXiv e-prints, arXiv:1901.05489.
\newblock \doarXiv{1901.05489}

\bibitem[{{Chollet, ~F. and Keras Team}(2015)}]{keras}
{Chollet, ~F. and Keras Team}. 2015, Keras

\bibitem[{{Conselice}(2018)}]{conselice}
{Conselice}, C.~J. 2018, Research Notes of the American Astronomical Society,
  2, 43, \dodoi{10.3847/2515-5172/aab7f6}

\bibitem[{{Dalcanton} {et~al.}(1997){Dalcanton}, {Spergel}, {Gunn}, {Schmidt},
  \& {Schneider}}]{dalcanton}
{Dalcanton}, J.~J., {Spergel}, D.~N., {Gunn}, J.~E., {Schmidt}, M., \&
  {Schneider}, D.~P. 1997, \aj, 114, 635, \dodoi{10.1086/118499}

\bibitem[{{Dey} {et~al.}(2019){Dey}, {Schlegel}, {Lang}, {Blum}, {Burleigh},
  {Fan}, {Findlay}, {Finkbeiner}, {Herrera}, {Juneau}, {Landriau}, {Levi},
  {McGreer}, {Meisner}, {Myers}, {Moustakas}, {Nugent}, {Patej}, {Schlafly},
  {Walker}, {Valdes}, {Weaver}, {Y{\`e}che}, {Zou}, {Zhou}, {Abareshi},
  {Abbott}, {Abolfathi}, {Aguilera}, {Alam}, {Allen}, {Alvarez}, {Annis},
  {Ansarinejad}, {Aubert}, {Beechert}, {Bell}, {BenZvi}, {Beutler}, {Bielby},
  {Bolton}, {Brice{\~n}o}, {Buckley-Geer}, {Butler}, {Calamida}, {Carlberg},
  {Carter}, {Casas}, {Castander}, {Choi}, {Comparat}, {Cukanovaite}, {Delubac},
  {DeVries}, {Dey}, {Dhungana}, {Dickinson}, {Ding}, {Donaldson}, {Duan},
  {Duckworth}, {Eftekharzadeh}, {Eisenstein}, {Etourneau}, {Fagrelius},
  {Farihi}, {Fitzpatrick}, {Font-Ribera}, {Fulmer}, {G{\"a}nsicke},
  {Gaztanaga}, {George}, {Gerdes}, {Gontcho}, {Gorgoni}, {Green}, {Guy},
  {Harmer}, {Hernand ez}, {Honscheid}, {Huang}, {James}, {Jannuzi}, {Jiang},
  {Joyce}, {Karcher}, {Karkar}, {Kehoe}, {Kneib}, {Kueter-Young}, {Lan},
  {Lauer}, {Le Guillou}, {Le Van Suu}, {Lee}, {Lesser}, {Perreault Levasseur},
  {Li}, {Mann}, {Marshall}, {Mart{\'\i}nez-V{\'a}zquez}, {Martini}, {du Mas des
  Bourboux}, {McManus}, {Meier}, {M{\'e}nard}, {Metcalfe},
  {Mu{\~n}oz-Guti{\'e}rrez}, {Najita}, {Napier}, {Narayan}, {Newman}, {Nie},
  {Nord}, {Norman}, {Olsen}, {Paat}, {Palanque-Delabrouille}, {Peng},
  {Poppett}, {Poremba}, {Prakash}, {Rabinowitz}, {Raichoor}, {Rezaie},
  {Robertson}, {Roe}, {Ross}, {Ross}, {Rudnick}, {Safonova}, {Saha},
  {S{\'a}nchez}, {Savary}, {Schweiker}, {Scott}, {Seo}, {Shan}, {Silva},
  {Slepian}, {Soto}, {Sprayberry}, {Staten}, {Stillman}, {Stupak}, {Summers},
  {Sien Tie}, {Tirado}, {Vargas-Maga{\~n}a}, {Vivas}, {Wechsler}, {Williams},
  {Yang}, {Yang}, {Yapici}, {Zaritsky}, {Zenteno}, {Zhang}, {Zhang}, {Zhou}, \&
  {Zhou}}]{dey}
{Dey}, A., {Schlegel}, D.~J., {Lang}, D., {et~al.} 2019, \aj, 157, 168,
  \dodoi{10.3847/1538-3881/ab089d}

\bibitem[{{Di Cintio} {et~al.}(2017){Di Cintio}, {Brook}, {Dutton},
  {Macci{\`o}}, {Obreja}, \& {Dekel}}]{cintio2017}
{Di Cintio}, A., {Brook}, C.~B., {Dutton}, A.~A., {et~al.} 2017, \mnras, 466,
  L1, \dodoi{10.1093/mnrasl/slw210}

\bibitem[{{Disney}(1976)}]{disney}
{Disney}, M.~J. 1976, \nat, 263, 573, \dodoi{10.1038/263573a0}

\bibitem[{{Drlica-Wagner} {et~al.}(2015){Drlica-Wagner}, {Bechtol}, {Rykoff},
  {Luque}, {Queiroz}, {Mao}, {Wechsler}, {Simon}, {Santiago}, {Yanny},
  {Balbinot}, {Dodelson}, {Fausti Neto}, {James}, {Li}, {Maia}, {Marshall},
  {Pieres}, {Stringer}, {Walker}, {Abbott}, {Abdalla}, {Allam},
  {Benoit-L{\'e}vy}, {Bernstein}, {Bertin}, {Brooks}, {Buckley-Geer}, {Burke},
  {Carnero Rosell}, {Carrasco Kind}, {Carretero}, {Crocce}, {da Costa},
  {Desai}, {Diehl}, {Dietrich}, {Doel}, {Eifler}, {Evrard}, {Finley},
  {Flaugher}, {Fosalba}, {Frieman}, {Gaztanaga}, {Gerdes}, {Gruen}, {Gruendl},
  {Gutierrez}, {Honscheid}, {Kuehn}, {Kuropatkin}, {Lahav}, {Martini},
  {Miquel}, {Nord}, {Ogando}, {Plazas}, {Reil}, {Roodman}, {Sako}, {Sanchez},
  {Scarpine}, {Schubnell}, {Sevilla-Noarbe}, {Smith}, {Soares-Santos},
  {Sobreira}, {Suchyta}, {Swanson}, {Tarle}, {Tucker}, {Vikram}, {Wester},
  {Zhang}, {Zuntz}, \& {DES Collaboration}}]{drlica-wagner}
{Drlica-Wagner}, A., {Bechtol}, K., {Rykoff}, E.~S., {et~al.} 2015, \apj, 813,
  109, \dodoi{10.1088/0004-637X/813/2/109}

\bibitem[{{Duc} {et~al.}(2015){Duc}, {Cuillandre}, {Karabal}, {Cappellari},
  {Alatalo}, {Blitz}, {Bournaud}, {Bureau}, {Crocker}, {Davies}, {Davis}, {de
  Zeeuw}, {Emsellem}, {Khochfar}, {Krajnovi{\'c}}, {Kuntschner}, {McDermid},
  {Michel-Dansac}, {Morganti}, {Naab}, {Oosterloo}, {Paudel}, {Sarzi}, {Scott},
  {Serra}, {Weijmans}, \& {Young}}]{duc}
{Duc}, P.-A., {Cuillandre}, J.-C., {Karabal}, E., {et~al.} 2015, \mnras, 446,
  120, \dodoi{10.1093/mnras/stu2019}

\bibitem[{{Fliri} \& {Trujillo}(2016)}]{fliri}
{Fliri}, J., \& {Trujillo}, I. 2016, Monthly Notices of the Royal Astronomical
  Society, 456, 1359, \dodoi{10.1093/mnras/stv2686}

\bibitem[{{Geach} {et~al.}(2017){Geach}, {Lin}, {Makler}, {Kneib}, {Ross},
  {Wang}, {Hsieh}, {Leauthaud}, {Bundy}, {McCracken}, {Comparat}, {Caminha},
  {Hudelot}, {Lin}, {Van Waerbeke}, {Pereira}, \& {Mast}}]{nearIR}
{Geach}, J.~E., {Lin}, Y.~T., {Makler}, M., {et~al.} 2017, The Astrophysical
  Journal Supplement Series, 231, 7, \dodoi{10.3847/1538-4365/aa74b6}

\bibitem[{{Ginsburg} {et~al.}(2019){Ginsburg}, {Sip{\H{o}}cz}, {Brasseur},
  {Cowperthwaite}, {Craig}, {Deil}, {Guillochon}, {Guzman}, {Liedtke}, {Lian
  Lim}, {Lockhart}, {Mommert}, {Morris}, {Norman}, {Parikh}, {Persson},
  {Robitaille}, {Segovia}, {Singer}, {Tollerud}, {de Val-Borro}, {Valtchanov},
  {Woillez}, {Astroquery Collaboration}, \& {a subset of astropy
  Collaboration}}]{astroquery}
{Ginsburg}, A., {Sip{\H{o}}cz}, B.~M., {Brasseur}, C.~E., {et~al.} 2019, \aj,
  157, 98, \dodoi{10.3847/1538-3881/aafc33}

\bibitem[{{Greco} {et~al.}(2018){Greco}, {Greene}, {Strauss}, {Macarthur},
  {Flowers}, {Goulding}, {Huang}, {Kim}, {Komiyama}, {Leauthaud}, {Leisman},
  {Lupton}, {Sif{\'o}n}, \& {Wang}}]{greco}
{Greco}, J.~P., {Greene}, J.~E., {Strauss}, M.~A., {et~al.} 2018, \apj, 857,
  104, \dodoi{10.3847/1538-4357/aab842}

\bibitem[{{Green}(2018)}]{green}
{Green}, G. 2018, The Journal of Open Source Software, 3, 695,
  \dodoi{10.21105/joss.00695}

\bibitem[{Haussler {et~al.}(2007)Haussler, McIntosh, Barden, Bell, Rix, Borch,
  Beckwith, Caldwell, Heymans, Jahnke, Jogee, Koposov, Meisenheimer, Sanchez,
  Somerville, Wisotzki, \& Wolf}]{Haussler}
Haussler, B., McIntosh, D.~H., Barden, M., {et~al.} 2007, The Astrophysical
  Journal Supplement Series, 172, 615.
\newblock \url{https://doi.org/10.1086/518836}

\bibitem[{Hinshaw {et~al.}(2013)Hinshaw, Larson, Komatsu, Spergel, Bennett,
  Dunkley, Nolta, Halpern, Hill, Odegard, Page, Smith, Weiland, Gold, Jarosik,
  Kogut, Limon, Meyer, Tucker, Wollack, \& Wright}]{wmap9}
Hinshaw, G., Larson, D., Komatsu, E., {et~al.} 2013, The Astrophysical Journal
  Supplement Series, 208, 19, \dodoi{10.1088/0067-0049/208/2/19}

\bibitem[{{Hodge} {et~al.}(2011){Hodge}, {Becker}, {White}, {Richards}, \&
  {Zeimann}}]{hodge}
{Hodge}, J.~A., {Becker}, R.~H., {White}, R.~L., {Richards}, G.~T., \&
  {Zeimann}, G.~R. 2011, The Astronomical Journal, 142, 3,
  \dodoi{10.1088/0004-6256/142/1/3}

\bibitem[{{Huang} {et~al.}(2017){Huang}, {Liu}, {Van Der Maaten}, \&
  {Weinberger}}]{densenet}
{Huang}, G., {Liu}, Z., {Van Der Maaten}, L., \& {Weinberger}, K.~Q. 2017, in
  2017 IEEE Conference on Computer Vision and Pattern Recognition (CVPR),
  2261--2269, \dodoi{10.1109/CVPR.2017.243}

\bibitem[{{Hunter}(2007)}]{matplotlib}
{Hunter}, J.~D. 2007, Computing in Science and Engineering, 9, 90,
  \dodoi{10.1109/MCSE.2007.55}

\bibitem[{{Impey} {et~al.}(1988){Impey}, {Bothun}, \& {Malin}}]{impey}
{Impey}, C., {Bothun}, G., \& {Malin}, D. 1988, \apj, 330, 634,
  \dodoi{10.1086/166500}

\bibitem[{{Impey} {et~al.}(1996){Impey}, {Sprayberry}, {Irwin}, \&
  {Bothun}}]{impey96}
{Impey}, C.~D., {Sprayberry}, D., {Irwin}, M.~J., \& {Bothun}, G.~D. 1996,
  \apjs, 105, 209, \dodoi{10.1086/192313}

\bibitem[{{Jiang} {et~al.}(2014){Jiang}, {Fan}, {Bian}, {McGreer}, {Strauss},
  {Annis}, {Buck}, {Green}, {Hodge}, {Myers}, {Rafiee}, \& {Richards}}]{Jiang}
{Jiang}, L., {Fan}, X., {Bian}, F., {et~al.} 2014, \apjs, 213, 12,
  \dodoi{10.1088/0067-0049/213/1/12}

\bibitem[{Jones {et~al.}(2001)Jones, Oliphant, P., \& et~al.}]{jones}
Jones, E., Oliphant, T., P., P., \& et~al. 2001, {SciPy}: Open source

\bibitem[{Kadowaki {et~al.}(2021)Kadowaki, Zaritsky, Donnerstein, RS,
  Karunakaran, \& Spekkens}]{kadowaki21}
Kadowaki, J., Zaritsky, D., Donnerstein, R., {et~al.} 2021

\bibitem[{{Kadowaki} {et~al.}(2017){Kadowaki}, {Zaritsky}, \&
  {Donnerstein}}]{kadowaki17}
{Kadowaki}, J., {Zaritsky}, D., \& {Donnerstein}, R.~L. 2017, \apjl, 838, L21,
  \dodoi{10.3847/2041-8213/aa653d}

\bibitem[{{Karunakaran} {et~al.}(2020){Karunakaran}, {Spekkens}, {Zaritsky},
  {Donnerstein}, {Kadowaki}, \& {Dey}}]{karunakaran}
{Karunakaran}, A., {Spekkens}, K., {Zaritsky}, D., {et~al.} 2020, \apj, 902,
  39, \dodoi{10.3847/1538-4357/abb464}

\bibitem[{Kingma \& Ba(2017)}]{king}
Kingma, D.~P., \& Ba, J. 2017, Adam: A Method for Stochastic Optimization.
\newblock \doarXiv{1412.6980}

\bibitem[{{Koda} {et~al.}(2015){Koda}, {Yagi}, {Yamanoi}, \& {Komiyama}}]{koda}
{Koda}, J., {Yagi}, M., {Yamanoi}, H., \& {Komiyama}, Y. 2015, \apjl, 807, L2,
  \dodoi{10.1088/2041-8205/807/1/L2}

\bibitem[{{Leisman} {et~al.}(2017){Leisman}, {Haynes}, {Janowiecki},
  {Hallenbeck}, {J{\'o}zsa}, {Giovanelli}, {Adams}, {Bernal Neira}, {Cannon},
  {Janesh}, {Rhode}, \& {Salzer}}]{leisman}
{Leisman}, L., {Haynes}, M.~P., {Janowiecki}, S., {et~al.} 2017, \apj, 842,
  133, \dodoi{10.3847/1538-4357/aa7575}

\bibitem[{{Lim} {et~al.}(2020){Lim}, {C{\^o}t{\'e}}, {Peng}, {Ferrarese},
  {Roediger}, {Durrell}, {Mihos}, {Wang}, {Gwyn}, {Cuillandre}, {Liu},
  {S{\'a}nchez-Janssen}, {Toloba}, {Sales}, {Guhathakurta}, {Lan{\c{c}}on}, \&
  {Puzia}}]{lim}
{Lim}, S., {C{\^o}t{\'e}}, P., {Peng}, E.~W., {et~al.} 2020, \apj, 899, 69,
  \dodoi{10.3847/1538-4357/aba433}

\bibitem[{{Makarov} {et~al.}(2015){Makarov}, {Sharina}, {Karachentseva}, \&
  {Karachentsev}}]{makarov}
{Makarov}, D.~I., {Sharina}, M.~E., {Karachentseva}, V.~E., \& {Karachentsev},
  I.~D. 2015, \aap, 581, A82, \dodoi{10.1051/0004-6361/201526947}

\bibitem[{{Mancera Pi{\~n}a} {et~al.}(2019){Mancera Pi{\~n}a}, {Aguerri},
  {Peletier}, {Venhola}, {Trager}, \& {Choque Challapa}}]{mancera}
{Mancera Pi{\~n}a}, P.~E., {Aguerri}, J.~A.~L., {Peletier}, R.~F., {et~al.}
  2019, \mnras, 485, 1036, \dodoi{10.1093/mnras/stz238}

\bibitem[{{Martin} {et~al.}(2019){Martin}, {Kaviraj}, {Laigle}, {Devriendt},
  {Jackson}, {Peirani}, {Dubois}, {Pichon}, \& {Slyz}}]{martin19}
{Martin}, G., {Kaviraj}, S., {Laigle}, C., {et~al.} 2019, \mnras, 485, 796,
  \dodoi{10.1093/mnras/stz356}

\bibitem[{{Mart{\'i}nez-Delgado} {et~al.}(2016){Mart{\'i}nez-Delgado},
  {L{\"a}sker}, {Sharina}, {Toloba}, {Fliri}, {Beaton}, {Valls-Gabaud},
  {Karachentsev}, {Chonis}, {Grebel}, {Forbes}, {Romanowsky},
  {Gallego-Laborda}, {Teuwen}, {G{\'o}mez-Flechoso}, {Wang}, {Guhathakurta},
  {Kaisin}, \& {Ho}}]{martinez}
{Mart{\'i}nez-Delgado}, D., {L{\"a}sker}, R., {Sharina}, M., {et~al.} 2016,
  \aj, 151, 96, \dodoi{10.3847/0004-6256/151/4/96}

\bibitem[{{McKinney}(2010)}]{pandas}
{McKinney}, W. 2010, Proceedings of the 9th Python in Science Conference, 51

\bibitem[{{Meisner} \& {Finkbeiner}(2014)}]{meisner}
{Meisner}, A.~M., \& {Finkbeiner}, D.~P. 2014, \apj, 781, 5,
  \dodoi{10.1088/0004-637X/781/1/5}

\bibitem[{{Mendes de Oliveira} {et~al.}(2019){Mendes de Oliveira}, {Ribeiro},
  {Schoenell}, {Kanaan}, {Overzier}, {Molino}, {Sampedro}, {Coelho}, {Barbosa},
  {Cortesi}, {Costa-Duarte}, {Herpich}, {Hernand ez-Jimenez}, {Placco},
  {Xavier}, {Abramo}, {Saito}, {Chies-Santos}, {Ederoclite}, {Lopes de
  Oliveira}, {Gon{\c{c}}alves}, {Akras}, {Almeida}, {Almeida-Fernandes},
  {Beers}, {Bonatto}, {Bonoli}, {Cypriano}, {de Lima}, {de Souza}, {de Souza},
  {Ferrari}, {Gon{\c{c}}alves}, {Gonzalez}, {Guti{\'e}rrez-Soto}, {Hartmann},
  {Jaffe}, {Kerber}, {Lima-Dias}, {Lopes}, {Menendez-Delmestre}, {Nakazono},
  {Novais}, {Ortega-Minakata}, {Pereira}, {Perottoni}, {Queiroz}, {Reis},
  {Santos}, {Santos-Silva}, {Santucci}, {Barbosa}, {Siffert}, {Sodr{\'e}},
  {Torres-Flores}, {Westera}, {Whitten}, {Alcaniz}, {Alonso-Garc{\'\i}a},
  {Alencar}, {Alvarez-Cand al}, {Amram}, {Azanha}, {Barb{\'a}},
  {Bernardinelli}, {Borges Fernandes}, {Branco}, {Brito-Silva}, {Buzzo},
  {Caffer}, {Campillay}, {Cano}, {Carvano}, {Castejon}, {Cid Fernandes},
  {Dantas}, {Daflon}, {Damke}, {de la Reza}, {de Azevedo}, {De Paula}, {Diem},
  {Donnerstein}, {Dors}, {Dupke}, {Eikenberry}, {Escudero}, {Faifer},
  {Far{\'\i}as}, {Fernandes}, {Fernandes}, {Fontes}, {Galarza}, {Hirata},
  {Katena}, {Gregorio-Hetem}, {Hern{\'a}ndez-Fern{\'a}ndez}, {Izzo},
  {Arancibia}, {Jatenco-Pereira}, {Jim{\'e}nez-Teja}, {Kann}, {Krabbe},
  {Labayru}, {Lazzaro}, {Lima Neto}, {Lopes}, {Magalh{\~a}es}, {Makler}, {de
  Menezes}, {Miralda-Escud{\'e}}, {Monteiro-Oliveira}, {Montero-Dorta},
  {Mu{\~n}oz-Elgueta}, {Nemmen}, {Castell{\'o}n}, {Oliveira}, {Ort{\'\i}z},
  {Pattaro}, {Pereira}, {Quint}, {Riguccini}, {Rocha Pinto}, {Rodrigues},
  {Roig}, {Rossi}, {Saha}, {Santos}, {Schnorr M{\"u}ller}, {Sesto}, {Silva},
  {Smith Castelli}, {Teixeira}, {Telles}, {Thom de Souza}, {Th{\"o}ne},
  {Trevisan}, {de Ugarte Postigo}, {Urrutia-Viscarra}, {Veiga}, {Vika},
  {Vitorelli}, {Werle}, {Werner}, \& {Zaritsky}}]{splus}
{Mendes de Oliveira}, C., {Ribeiro}, T., {Schoenell}, W., {et~al.} 2019,
  Monthly Notices of the Royal Astronomical Society, 2048,
  \dodoi{10.1093/mnras/stz1985}

\bibitem[{{Mihos} {et~al.}(2015){Mihos}, {Durrell}, {Ferrarese}, {Feldmeier},
  {C{\^o}t{\'e}}, {Peng}, {Harding}, {Liu}, {Gwyn}, \& {Cuillandre}}]{mihos}
{Mihos}, J.~C., {Durrell}, P.~R., {Ferrarese}, L., {et~al.} 2015, \apjl, 809,
  L21, \dodoi{10.1088/2041-8205/809/2/L21}

\bibitem[{{Millman} \& {Aivazis}(2011)}]{scipy2}
{Millman}, K.~J., \& {Aivazis}, M. 2011, Computing in Science and Engineering,
  13, 9, \dodoi{10.1109/MCSE.2011.36}

\bibitem[{{Mu{\~n}oz} {et~al.}(2015){Mu{\~n}oz}, {Eigenthaler}, {Puzia},
  {Taylor}, {Ordenes-Brice{\~n}o}, {Alamo-Mart{\'\i}nez}, {Ribbeck},
  {{\'A}ngel}, {Capaccioli}, {C{\^o}t{\'e}}, {Ferrarese}, {Galaz}, {Hempel},
  {Hilker}, {Jord{\'a}n}, {Lan{\c{c}}on}, {Mieske}, {Paolillo}, {Richtler},
  {S{\'a}nchez-Janssen}, \& {Zhang}}]{munoz}
{Mu{\~n}oz}, R.~P., {Eigenthaler}, P., {Puzia}, T.~H., {et~al.} 2015, \apjl,
  813, L15, \dodoi{10.1088/2041-8205/813/1/L15}

\bibitem[{Newville {et~al.}(2014)Newville, Stensitzki, Allen, \&
  Ingargiola}]{newville}
Newville, M., Stensitzki, T., Allen, D.~B., \& Ingargiola, A. 2014, {LMFIT:
  Non-Linear Least-Square Minimization and Curve-Fitting for Python}

\bibitem[{{Oke}(1964)}]{oke1}
{Oke}, J.~B. 1964, \apj, 140, 689, \dodoi{10.1086/147960}

\bibitem[{{Oke} \& {Gunn}(1983)}]{oke2}
{Oke}, J.~B., \& {Gunn}, J.~E. 1983, \apj, 266, 713, \dodoi{10.1086/160817}

\bibitem[{{Oliphant}(2007)}]{scipy1}
{Oliphant}, T.~E. 2007, Computing in Science and Engineering, 9, 10,
  \dodoi{10.1109/MCSE.2007.58}

\bibitem[{Pedregosa {et~al.}(2011)Pedregosa, Varoquaux, Gramfort, Michel,
  Thirion, Grisel, Blondel, Prettenhofer, Weiss, Dubourg, Vanderplas, Passos,
  Cournapeau, Brucher, Perrot, \& Duchesnay}]{sklearn}
Pedregosa, F., Varoquaux, G., Gramfort, A., {et~al.} 2011, Journal of Machine
  Learning Research, 12, 2825

\bibitem[{{Peng} {et~al.}(2002){Peng}, {Ho}, {Impey}, \& {Rix}}]{peng}
{Peng}, C.~Y., {Ho}, L.~C., {Impey}, C.~D., \& {Rix}, H.-W. 2002, \aj, 124,
  266, \dodoi{10.1086/340952}

\bibitem[{{Penny} {et~al.}(2009){Penny}, {Conselice}, {De Rijcke}, \&
  {Held}}]{penny}
{Penny}, S.~J., {Conselice}, C.~J., {De Rijcke}, S., \& {Held}, E.~V. 2009,
  Astronomische Nachrichten, 330, 991, \dodoi{10.1002/asna.200911276}

\bibitem[{{Planck Collaboration} {et~al.}(2014){Planck Collaboration},
  {Abergel}, {Ade}, {Aghanim}, {Alves}, {Aniano}, {Armitage-Caplan}, {Arnaud},
  {Ashdown}, {Atrio-Barand ela}, {Aumont}, {Baccigalupi}, {Banday}, {Barreiro},
  {Bartlett}, {Battaner}, {Benabed}, {Beno{\^\i}t}, {Benoit-L{\'e}vy},
  {Bernard}, {Bersanelli}, {Bielewicz}, {Bobin}, {Bock}, {Bonaldi}, {Bond},
  {Borrill}, {Bouchet}, {Boulanger}, {Bridges}, {Bucher}, {Burigana}, {Butler},
  {Cardoso}, {Catalano}, {Chamballu}, {Chary}, {Chiang}, {Chiang},
  {Christensen}, {Church}, {Clemens}, {Clements}, {Colombi}, {Colombo},
  {Combet}, {Couchot}, {Coulais}, {Crill}, {Curto}, {Cuttaia}, {Danese},
  {Davies}, {Davis}, {de Bernardis}, {de Rosa}, {de Zotti}, {Delabrouille},
  {Delouis}, {D{\'e}sert}, {Dickinson}, {Diego}, {Dole}, {Donzelli},
  {Dor{\'e}}, {Douspis}, {Draine}, {Dupac}, {Efstathiou}, {En{\ss}lin},
  {Eriksen}, {Falgarone}, {Finelli}, {Forni}, {Frailis}, {Fraisse},
  {Franceschi}, {Galeotta}, {Ganga}, {Ghosh}, {Giard}, {Giardino},
  {Giraud-H{\'e}raud}, {Gonz{\'a}lez-Nuevo}, {G{\'o}rski}, {Gratton},
  {Gregorio}, {Grenier}, {Gruppuso}, {Guillet}, {Hansen}, {Hanson}, {Harrison},
  {Helou}, {Henrot-Versill{\'e}}, {Hern{\'a}ndez-Monteagudo}, {Herranz},
  {Hildebrand t}, {Hivon}, {Hobson}, {Holmes}, {Hornstrup}, {Hovest},
  {Huffenberger}, {Jaffe}, {Jaffe}, {Jewell}, {Joncas}, {Jones}, {Juvela},
  {Keih{\"a}nen}, {Keskitalo}, {Kisner}, {Knoche}, {Knox}, {Kunz},
  {Kurki-Suonio}, {Lagache}, {L{\"a}hteenm{\"a}ki}, {Lamarre}, {Lasenby},
  {Laureijs}, {Lawrence}, {Leonardi}, {Le{\'o}n-Tavares}, {Lesgourgues},
  {Levrier}, {Liguori}, {Lilje}, {Linden-V{\o}rnle}, {L{\'o}pez-Caniego},
  {Lubin}, {Mac{\'\i}as-P{\'e}rez}, {Maffei}, {Maino}, {Mand olesi}, {Maris},
  {Marshall}, {Martin}, {Mart{\'\i}nez-Gonz{\'a}lez}, {Masi}, {Massardi},
  {Matarrese}, {Matthai}, {Mazzotta}, {McGehee}, {Melchiorri}, {Mendes},
  {Mennella}, {Migliaccio}, {Mitra}, {Miville-Desch{\^e}nes}, {Moneti},
  {Montier}, {Morgante}, {Mortlock}, {Munshi}, {Murphy}, {Naselsky}, {Nati},
  {Natoli}, {Netterfield}, {N{\o}rgaard-Nielsen}, {Noviello}, {Novikov},
  {Novikov}, {Osborne}, {Oxborrow}, {Paci}, {Pagano}, {Pajot}, {Paladini},
  {Paoletti}, {Pasian}, {Patanchon}, {Perdereau}, {Perotto}, {Perrotta},
  {Piacentini}, {Piat}, {Pierpaoli}, {Pietrobon}, {Plaszczynski},
  {Pointecouteau}, {Polenta}, {Ponthieu}, {Popa}, {Poutanen}, {Pratt},
  {Pr{\'e}zeau}, {Prunet}, {Puget}, {Rachen}, {Reach}, {Rebolo}, {Reinecke},
  {Remazeilles}, {Renault}, {Ricciardi}, {Riller}, {Ristorcelli}, {Rocha},
  {Rosset}, {Roudier}, {Rowan-Robinson}, {Rubi{\~n}o-Mart{\'\i}n}, {Rusholme},
  {Sandri}, {Santos}, {Savini}, {Scott}, {Seiffert}, {Shellard}, {Spencer},
  {Starck}, {Stolyarov}, {Stompor}, {Sudiwala}, {Sunyaev}, {Sureau}, {Sutton},
  {Suur-Uski}, {Sygnet}, {Tauber}, {Tavagnacco}, {Terenzi}, {Toffolatti},
  {Tomasi}, {Tristram}, {Tucci}, {Tuovinen}, {T{\"u}rler}, {Umana},
  {Valenziano}, {Valiviita}, {Van Tent}, {Verstraete}, {Vielva}, {Villa},
  {Vittorio}, {Wade}, {Wandelt}, {Welikala}, {Ysard}, {Yvon}, {Zacchei}, \&
  {Zonca}}]{planck}
{Planck Collaboration}, {Abergel}, A., {Ade}, P.~A.~R., {et~al.} 2014, \aap,
  571, A11, \dodoi{10.1051/0004-6361/201323195}

\bibitem[{{Prole} {et~al.}(2018){Prole}, {Davies}, {Keenan}, \&
  {Davies}}]{prole}
{Prole}, D.~J., {Davies}, J.~I., {Keenan}, O.~C., \& {Davies}, L.~J.~M. 2018,
  \mnras, 478, 667, \dodoi{10.1093/mnras/sty1021}

\bibitem[{{Rom{\'a}n} \& {Trujillo}(2017{\natexlab{a}})}]{roman17a}
{Rom{\'a}n}, J., \& {Trujillo}, I. 2017{\natexlab{a}}, \mnras, 468, 703,
  \dodoi{10.1093/mnras/stx438}

\bibitem[{{Rom{\'a}n} \& {Trujillo}(2017{\natexlab{b}})}]{roman17b}
---. 2017{\natexlab{b}}, \mnras, 468, 4039, \dodoi{10.1093/mnras/stx694}

\bibitem[{{Rom{\'a}n} \& {Trujillo}(2018)}]{roman18}
---. 2018, Research Notes of the American Astronomical Society, 2, 144,
  \dodoi{10.3847/2515-5172/aad8b8}

\bibitem[{{Rom{\'a}n} {et~al.}(2019){Rom{\'a}n}, {Trujillo}, \&
  {Montes}}]{roman19}
{Rom{\'a}n}, J., {Trujillo}, I., \& {Montes}, M. 2019, arXiv e-prints,
  arXiv:1907.00978.
\newblock \doarXiv{1907.00978}

\bibitem[{{Sales} {et~al.}(2020){Sales}, {Navarro}, {Pe{\~n}afiel}, {Peng},
  {Lim}, \& {Hernquist}}]{sales20}
{Sales}, L.~V., {Navarro}, J.~F., {Pe{\~n}afiel}, L., {et~al.} 2020, \mnras,
  494, 1848, \dodoi{10.1093/mnras/staa854}

\bibitem[{{Sandage} \& {Binggeli}(1984)}]{sandage}
{Sandage}, A., \& {Binggeli}, B. 1984, \aj, 89, 919, \dodoi{10.1086/113588}

\bibitem[{Schlafly \& Finkbeiner(2011)}]{Schlafly}
Schlafly, E.~F., \& Finkbeiner, D.~P. 2011, The Astrophysical Journal, 737,
  103, \dodoi{10.1088/0004-637x/737/2/103}

\bibitem[{{Schlegel} {et~al.}(2011){Schlegel}, {Abdalla}, {Abraham}, {Ahn},
  {Allende Prieto}, {Annis}, {Aubourg}, {Azzaro}, {Baltay}, {Baugh}, {Bebek},
  {Becerril}, {Blanton}, {Bolton}, {Bromley}, {Cahn}, {Carton},
  {Cervantes-Cota}, {Chu}, {Cortes}, {Dawson}, {Dey}, {Dickinson}, {Diehl},
  {Doel}, {Ealet}, {Edelstein}, {Eppelle}, {Escoffier}, {Evrard}, {Faccioli},
  {Frenk}, {Geha}, {Gerdes}, {Gondolo}, {Gonzalez-Arroyo}, {Grossan},
  {Heckman}, {Heetderks}, {Ho}, {Honscheid}, {Huterer}, {Ilbert}, {Ivans},
  {Jelinsky}, {Jing}, {Joyce}, {Kennedy}, {Kent}, {Kieda}, {Kim}, {Kim},
  {Kneib}, {Kong}, {Kosowsky}, {Krishnan}, {Lahav}, {Lampton}, {LeBohec}, {Le
  Brun}, {Levi}, {Li}, {Liang}, {Lim}, {Lin}, {Linder}, {Lorenzon}, {de la
  Macorra}, {Magneville}, {Malina}, {Marinoni}, {Martinez}, {Majewski},
  {Matheson}, {McCloskey}, {McDonald}, {McKay}, {McMahon}, {Menard},
  {Miralda-Escude}, {Modjaz}, {Montero-Dorta}, {Morales}, {Mostek}, {Newman},
  {Nichol}, {Nugent}, {Olsen}, {Padmanabhan}, {Palanque-Delabrouille}, {Park},
  {Peacock}, {Percival}, {Perlmutter}, {Peroux}, {Petitjean}, {Prada},
  {Prieto}, {Prochaska}, {Reil}, {Rockosi}, {Roe}, {Rollinde}, {Roodman},
  {Ross}, {Rudnick}, {Ruhlmann-Kleider}, {Sanchez}, {Sawyer}, {Schimd},
  {Schubnell}, {Scoccimaro}, {Seljak}, {Seo}, {Sheldon}, {Sholl},
  {Shulte-Ladbeck}, {Slosar}, {Smith}, {Smoot}, {Springer}, {Stril}, {Szalay},
  {Tao}, {Tarle}, {Taylor}, {Tilquin}, {Tinker}, {Valdes}, {Wang}, {Wang},
  {Weaver}, {Weinberg}, {White}, {Wood-Vasey}, {Yang}, {Yeche}, {Zakamska},
  {Zentner}, {Zhai}, \& {Zhang}}]{schlegel}
{Schlegel}, D., {Abdalla}, F., {Abraham}, T., {et~al.} 2011, arXiv e-prints,
  arXiv:1106.1706.
\newblock \doarXiv{1106.1706}

\bibitem[{{Schlegel} {et~al.}(1998){Schlegel}, {Finkbeiner}, \& {Davis}}]{SFD}
{Schlegel}, D.~J., {Finkbeiner}, D.~P., \& {Davis}, M. 1998, \apj, 500, 525,
  \dodoi{10.1086/305772}

\bibitem[{{Schombert} \& {Bothun}(1988)}]{schombert}
{Schombert}, J.~M., \& {Bothun}, G.~D. 1988, \aj, 95, 1389,
  \dodoi{10.1086/114736}

\bibitem[{{Schwartzenberg} {et~al.}(1995){Schwartzenberg}, {Phillipps},
  {Smith}, {Couch}, \& {Boyle}}]{sch}
{Schwartzenberg}, J.~M., {Phillipps}, S., {Smith}, R.~M., {Couch}, W.~J., \&
  {Boyle}, B.~J. 1995, \mnras, 275, 121, \dodoi{10.1093/mnras/275.1.121}

\bibitem[{{Shi} {et~al.}(2017){Shi}, {Zheng}, {Zhao}, {Pan}, {Li}, {Zou},
  {Zhou}, {Guo}, {An}, \& {Li}}]{shi}
{Shi}, D.~D., {Zheng}, X.~Z., {Zhao}, H.~B., {et~al.} 2017, \apj, 846, 26,
  \dodoi{10.3847/1538-4357/aa8327}

\bibitem[{{Singh} {et~al.}(2019){Singh}, {Zaritsky}, {Donnerstein}, \&
  {Spekkens}}]{rs}
{Singh}, P.~R., {Zaritsky}, D., {Donnerstein}, R., \& {Spekkens}, K. 2019, \aj,
  157, 212, \dodoi{10.3847/1538-3881/ab16f2}

\bibitem[{{Sprayberry} {et~al.}(1997){Sprayberry}, {Impey}, {Irwin}, \&
  {Bothun}}]{sprayberry}
{Sprayberry}, D., {Impey}, C.~D., {Irwin}, M.~J., \& {Bothun}, G.~D. 1997,
  \apj, 482, 104, \dodoi{10.1086/304126}

\bibitem[{{Steidel} {et~al.}(2000){Steidel}, {Adelberger}, {Shapley},
  {Pettini}, {Dickinson}, \& {Giavalisco}}]{steidel}
{Steidel}, C.~C., {Adelberger}, K.~L., {Shapley}, A.~E., {et~al.} 2000, \apj,
  532, 170, \dodoi{10.1086/308568}

\bibitem[{{Takey} {et~al.}(2016){Takey}, {Durret}, {Mahmoud}, \& {Ali}}]{xmm}
{Takey}, A., {Durret}, F., {Mahmoud}, E., \& {Ali}, G.~B. 2016, Astronomy and
  Astrophysics, 594, A32, \dodoi{10.1051/0004-6361/201628105}

\bibitem[{Tan \& Le(2020)}]{efficientnet}
Tan, M., \& Le, Q.~V. 2020, EfficientNet: Rethinking Model Scaling for
  Convolutional Neural Networks.
\newblock \doarXiv{1905.11946}

\bibitem[{{Tanoglidis} {et~al.}(2021){Tanoglidis}, {Drlica-Wagner}, {Wei},
  {Li}, {S{\'a}nchez}, {Zhang}, {Peter}, {Feldmeier-Krause}, {Prat}, {Casey},
  {Palmese}, {S{\'a}nchez}, {DeRose}, {Conselice}, {Gagnon}, {Abbott},
  {Aguena}, {Allam}, {Avila}, {Bechtol}, {Bertin}, {Bhargava}, {Brooks},
  {Burke}, {Rosell}, {Kind}, {Carretero}, {Chang}, {Costanzi}, {da Costa}, {De
  Vicente}, {Desai}, {Diehl}, {Doel}, {Eifler}, {Everett}, {Evrard},
  {Flaugher}, {Frieman}, {Garc{\'\i}a-Bellido}, {Gerdes}, {Gruendl},
  {Gschwend}, {Gutierrez}, {Hartley}, {Hollowood}, {Huterer}, {James},
  {Krause}, {Kuehn}, {Kuropatkin}, {Maia}, {March}, {Marshall}, {Menanteau},
  {Miquel}, {Ogando}, {Paz-Chinch{\'o}n}, {Romer}, {Roodman}, {Sanchez},
  {Scarpine}, {Serrano}, {Sevilla-Noarbe}, {Smith}, {Suchyta}, {Tarle},
  {Thomas}, {Tucker}, {Walker}, \& {DES Collaboration}}]{tanoglidis}
{Tanoglidis}, D., {Drlica-Wagner}, A., {Wei}, K., {et~al.} 2021, \apjs, 252,
  18, \dodoi{10.3847/1538-4365/abca89}

\bibitem[{{The DESI Collaboration}(2016{\natexlab{a}})}]{desi1}
{The DESI Collaboration}. 2016{\natexlab{a}}

\bibitem[{{The DESI Collaboration}(2016{\natexlab{b}})}]{desi2}
---. 2016{\natexlab{b}}

\bibitem[{Valdes {et~al.}(2014)Valdes, Gruendl, \& {DES Project}}]{valdes}
Valdes, F., Gruendl, R., \& {DES Project}. 2014

\bibitem[{{van der Burg} {et~al.}(2017){van der Burg}, {Hoekstra}, {Muzzin},
  {Sif{\'o}n}, {Viola}, {Bremer}, {Brough}, {Driver}, {Erben}, {Heymans},
  {Hildebrandt}, {Holwerda}, {Klaes}, {Kuijken}, {McGee}, {Nakajima},
  {Napolitano}, {Norberg}, {Taylor}, \& {Valentijn}}]{vdb17}
{van der Burg}, R. F.~J., {Hoekstra}, H., {Muzzin}, A., {et~al.} 2017, \aap,
  607, A79, \dodoi{10.1051/0004-6361/201731335}

\bibitem[{{van der Walt} {et~al.}(2011){van der Walt}, {Colbert}, \&
  {Varoquaux}}]{numpy}
{van der Walt}, S., {Colbert}, S.~C., \& {Varoquaux}, G. 2011, Computing in
  Science and Engineering, 13, 22, \dodoi{10.1109/MCSE.2011.37}

\bibitem[{{van Dokkum} {et~al.}(2016){van Dokkum}, {Abraham}, {Brodie},
  {Conroy}, {Danieli}, {Merritt}, {Mowla}, {Romanowsky}, \& {Zhang}}]{vdk16}
{van Dokkum}, P., {Abraham}, R., {Brodie}, J., {et~al.} 2016, \apjl, 828, L6,
  \dodoi{10.3847/2041-8205/828/1/L6}

\bibitem[{{van Dokkum} {et~al.}(2018){van Dokkum}, {Danieli}, {Cohen},
  {Merritt}, {Romanowsky}, {Abraham}, {Brodie}, {Conroy}, {Lokhorst}, {Mowla},
  {O'Sullivan}, \& {Zhang}}]{vdk18}
{van Dokkum}, P., {Danieli}, S., {Cohen}, Y., {et~al.} 2018, \nat, 555, 629,
  \dodoi{10.1038/nature25767}

\bibitem[{{van Dokkum} {et~al.}(2019){van Dokkum}, {Wasserman}, {Danieli},
  {Abraham}, {Brodie}, {Conroy}, {Forbes}, {Martin}, {Matuszewski},
  {Romanowsky}, \& {Villaume}}]{vdk19}
{van Dokkum}, P., {Wasserman}, A., {Danieli}, S., {et~al.} 2019, \apj, 880, 91,
  \dodoi{10.3847/1538-4357/ab2914}

\bibitem[{{van Dokkum} {et~al.}(2015{\natexlab{a}}){van Dokkum}, {Abraham},
  {Merritt}, {Zhang}, {Geha}, \& {Conroy}}]{vdk15a}
{van Dokkum}, P.~G., {Abraham}, R., {Merritt}, A., {et~al.} 2015{\natexlab{a}},
  \apjl, 798, L45, \dodoi{10.1088/2041-8205/798/2/L45}

\bibitem[{{van Dokkum} {et~al.}(2015{\natexlab{b}}){van Dokkum}, {Romanowsky},
  {Abraham}, {Brodie}, {Conroy}, {Geha}, {Merritt}, {Villaume}, \&
  {Zhang}}]{vdk15b}
{van Dokkum}, P.~G., {Romanowsky}, A.~J., {Abraham}, R., {et~al.}
  2015{\natexlab{b}}, \apjl, 804, L26, \dodoi{10.1088/2041-8205/804/1/L26}

\bibitem[{{Venhola} {et~al.}(2017){Venhola}, {Peletier}, {Laurikainen}, {Salo},
  {Lisker}, {Iodice}, {Capaccioli}, {Verdois Kleijn}, {Valentijn}, {Mieske},
  {Hilker}, {Wittmann}, {van de Ven}, {Grado}, {Spavone}, {Cantiello},
  {Napolitano}, {Paolillo}, \& {Falc{\'o}n-Barroso}}]{venhola}
{Venhola}, A., {Peletier}, R., {Laurikainen}, E., {et~al.} 2017, \aap, 608,
  A142, \dodoi{10.1051/0004-6361/201730696}

\bibitem[{{Wittmann} {et~al.}(2017){Wittmann}, {Lisker}, {Ambachew Tilahun},
  {Grebel}, {Conselice}, {Penny}, {Janz}, {Gallagher}, {Kotulla}, \&
  {McCormac}}]{wittmann}
{Wittmann}, C., {Lisker}, T., {Ambachew Tilahun}, L., {et~al.} 2017, \mnras,
  470, 1512, \dodoi{10.1093/mnras/stx1229}

\bibitem[{{Zaritsky} {et~al.}(2008){Zaritsky}, {Zabludoff}, \&
  {Gonzalez}}]{z08}
{Zaritsky}, D., {Zabludoff}, A.~I., \& {Gonzalez}, A.~H. 2008, \apj, 682, 68,
  \dodoi{10.1086/529577}

\bibitem[{{Zaritsky} {et~al.}(2019){Zaritsky}, {Donnerstein}, {Dey},
  {Kadowaki}, {Zhang}, {Karunakaran}, {Mart{\'\i}nez-Delgado}, {Rahman}, \&
  {Spekkens}}]{smudges}
{Zaritsky}, D., {Donnerstein}, R., {Dey}, A., {et~al.} 2019, \apjs, 240, 1,
  \dodoi{10.3847/1538-4365/aaefe9}

\end{thebibliography}
\bibliographystyle{aasjournal}

\end{document}